\documentclass[12pt,preprint]{aastex}

\newcommand{\msun}{\mbox{$\:M_{\sun}$}}

\newcommand{\noprint}[1]{}
\newcommand{\figsetstart}{{\bf Fig. Set} }
\newcommand{\figsetend}{}
\newcommand{\figsetgrpstart}{}
\newcommand{\figsetgrpend}{}
\newcommand{\figsetnum}[1]{{\bf #1.}}
\newcommand{\figsettitle}[1]{ {\bf #1} }
\newcommand{\figsetgrpnum}[1]{\noprint{#1}}
\newcommand{\figsetgrptitle}[1]{\noprint{#1}}
\newcommand{\figsetplot}[1]{\noprint{#1}}
\newcommand{\figsetgrpnote}[1]{\noprint{#1}}

\begin{document}

\title{A Survey of \textit{Far Ultraviolet Spectroscopic Explorer} Observations of Cataclysmic Variables\footnote{Based on observations made with the NASA-CNES-CSA Far Ultraviolet Spectroscopic Explorer. \textit{FUSE} was operated for NASA by the Johns Hopkins University under NASA
contract NAS5-32985.}}

\author{Cynthia S.\ Froning}
\email{cynthia.froning@colorado.edu}
\affil{Center for Astrophysics and Space Astronomy, \\
University of Colorado, 593 UCB, Boulder, CO, 80309}

\author{Knox S.\ Long}
\email{long@stsci.edu} \affil{Space Telescope Science Institute,
\\ 3700 San Martin Drive, Baltimore, MD 21218}

\author{Boris G\"{a}nsicke}
\email{Boris.Gaensicke@warwick.ac.uk} \affil{Department of Physics, \\
  University of Warwick, Coventry CV4 7AL, UK}

\and

\author{Paula Szkody} \email{szkody@alicar.astro.washington.edu}
\affil{Department of Astronomy, \\ University of Washington, Box 351580,
Seattle, WA 98195-1580}

\begin{abstract}
During its lifetime, the Far Ultraviolet Spectroscopic Explorer (\textit{FUSE}) was used to observe 99 cataclysmic variables in 211 separate observations. Here, we present a survey of the moderate resolution (R$\simeq$10,000), far-ultraviolet (905 -- 1188 \AA), time-averaged \textit{FUSE} spectra of cataclysmic variables (CVs).  The \textit{FUSE} spectra are morphologically diverse.  They show contributions from the accretion disk, the disk chromosphere, disk outflows, and the white dwarf, but the relative contribution of each component varies widely as a function of CV subtype, orbital period and evolutionary state, inclination, mass accretion rate, and magnetic field strength of the white dwarf. The data reveal information about the structure, temperature, density and mass flow rates of the disk and disk winds, the temperature of the white dwarf and the effects of ongoing accretion on its structure, and the long-term response of the systems to disk outbursts. The complete atlas of time-averaged \textit{FUSE} spectra of CVs are available at the Multimission Archive at Space Telescope Science Institute as a High Level Science Product.
\end{abstract}

\section{Introduction} \label{sec_intro}

The Far Ultraviolet Spectroscopic Explorer (\textit{FUSE}) was a NASA-supported astronomical observatory that operated from 1999 to 2007, acquiring far-ultraviolet (FUV; 905 -- 1188 \AA) spectra at moderate spectral resolution \citep[R$\simeq$10000][]{moos2000, sahnow2000}. \textit{FUSE} was used to observe a variety of Galactic and extragalactic targets, resulting in numerous scientific contributions in the areas of cosmology, galaxy structure, star formation and evolution, and planet formation.  

During its operational lifetime, \textit{FUSE} was used to observe 99 different cataclysmic variables (CVs) over a total of 211 observations.  CVs are mass-exchanging binary systems in which a late-type donor star transfers mass to a white dwarf (WD). In most CVs, mass transfer occurs via an accretion disk around the WD. The properties of CVs are discussed in detail by \citet{warner1995}. CVs are divided into a number of sub-classes.  First, they are distinguished between magnetic and non-magnetic systems. In magnetic CVs, the WD field strengths are large enough (B$\gtrsim$100~kG) to partially or completely disrupt the accretion disk. In non-magnetic CVs, the systems are divided between those systems that show semi-regular outbursts in the accretion disk (dwarf novae; DN) and those that do not (novalikes; NLs).   

CVs provide an excellent population of nearby, non-obscured targets in which to study the physics of mass accretion \citep{warner1995}.  Accretion disks are ubiquitous in astrophysical accretion, from compact binaries and proto-stars to the the central engines of active galaxies. CVs provide a way to probe disk accretion in multiple configurations and under a variety of stimuli.  The far-ultraviolet (FUV) in particular is an excellent window through which to view the accretion disk, disk outflows, and the WD accretor in CVs \citep[e.g.,][]{ladous1991,cordova1982,drew1985,gansicke2005}.  In high accretion rate systems (DN in outburst and NLs), the accretion disk can reach temperatures of up to $\simeq$60,000~K \citep{hartley2005} and the disk SED typically peaks in the FUV.  As a result, the FUV is a sensitive indicator of the inner disk structure and the mass accretion rate onto the WD. The FUV is also rich in spectral lines in a variety of species and ionization states that target the temperature and ionization structure of the disk, disk outflows, and the magnetically-channeled accretion column in magnetic CVs. 

In low state systems (quiescent DN and some magnetic CVs), the FUV continuum is often dominated by the WD.  The spectrum of the WD contains clues to the evolutionary and accretion history of the binary.  In the FUV, the wealth of spectral lines allows for determinations of metal abundances on the WD surface, which often show anomalies indicative of CNO processing of the accreted material, an artifact of the binary evolution in the system. Moreover, the physical properties of the interaction region between the accretion disk and the WD in CVs (the boundary layer) remains poorly understood \citep{ferland1982,godon2005,long2009}. At longer wavelengths it can be difficult to distinguish the spectrum of the WD from that of the disk, but observations at FUV wavelengths give the opportunity to determine the properties of the WD, the quiescent disk, and the boundary layer seperately.  Low-state CVs also show rich emission line spectra \citep{ladous1990}. The emission lines are usually assumed to originate in a chromosphere above the disk, but the location and properties of the line-emitting region and the vertical structure of the accretion disk are not well determined.  FUV spectroscopy places important constraints on the properties of the line-emitting region.

Most of the analysis of CVs in the FUV to date has proceeded on a case-by-case basis. While intensive studies of individual systems are necessary, it is also important to complement that work with a broad survey of the FUV properties of CVs \citep[e.g.,][]{verbunt1987,mauche1997, araujo2005, hamilton2007,sion2008}.  By discerning broad trends, we can place the behavior of individual systems in context and determine which characteristics of compact binaries determine their observational properties.  In quiescent systems, a survey of targets can facilitate separation of the WD component from the disk and chromospheric components and help to determine the structure and physical extent of the latter. For high state systems, observations of a wide range of systems allow us to determine the distribution of vertically extended material in accretion disks.  In all systems, a broad survey is necessary to analyze the relative FUV contributions of and trends in the properties of the accretion disk, winds, and the WD as a function of the viewing inclination, binary geometry, mass accretion rate, WD magnetic field strength, and evolutionary history.

To enable a broad survey of the FUV properties of CVs, we present a complete atlas of the \textit{FUSE} observations of CVs. This is the first large survey of the UV properties of CVs since the IUE atlas of \citet{ladous1990}, and the first to cover the FUV wavelengths below Ly$\alpha$. The time-averaged spectra of all the observations are available at the Multimission Archive at STScI (MAST) as a High Level Science Product. (\textit{FUSE} also observed a number of novae, supersoft X-ray sources, and X-ray binaries.  These are not included here.) In \S~\ref{sec_obs}, we summarize the observations and data calibration. In \S~\ref{sec_analysis}, we present the time-averaged spectra and examine the properties of CVs in the FUV.  Finally, we present concluding remarks in \S~\ref{sec_conclusion}. 

\section{\textit{FUSE} Observations and Calibration} \label{sec_obs}

The \textit{FUSE} telescope collected data simultaneously in four optical channels: LiF1, LiF2, SiC1, and SiC2. The optical elements in the LiF1 and LiF2 channels were coated with aluminum plus lithium fluoride to maximize sensitivity at wavelengths $>$1050~\AA\ while the SiC1 and SiC2 channels were coated with silicon carbide for best performance $<$1020~\AA. Each channel consisted of two non-contiguous segments  (e.g., LiF1A, LiF1B, etc.).  Together, the eight segments covered the overall wavelength range of the instrument (905--1187~\AA) with some overlap. The detectors consisted of two micro-channel plate devices, with the LiF1 and SiC1 spectra falling on one detector, and the LiF2 and SiC2 spectra on the other. \textit{FUSE} had three observing apertures.  Most of the observations made with \textit{FUSE} used the LWRS aperture ($30\arcsec \times 30 \arcsec$) although a few CVs were observed through the MDRS aperture ($4\arcsec \times 20 \arcsec$); these exceptions are noted in Tables~\ref{tab_list}.  The high resolution aperture (HIRS; $1.25\arcsec \times 20\arcsec$) was rarely used because of difficulties in maintaining co-alignment of the four optical channels to the required precision. The design and performance of \textit{FUSE} is presented in detail in the \textit{FUSE} Archival Instrument Handbook\footnote{http://archive.stsci.edu/fuse/ih.html}.

The default observation mode was the ``time-tag" observing mode, in which individual photon arrival events were recorded. The X and Y positions on the detector were recorded as well as the pulse height of each event.  A time stamp was inserted into the data stream every second, resulting in a relative accuracy of timing of 1 second.  Time-tag mode was the preferred observing mode because the resulting photon list could be screened to correct for bad time intervals, variable background, etc. The instrument data recorder could not store the full time-tag list when count rates exceeded 2500 counts sec$^{-1}$, however. For targets with higher count rates, the data were recorded in spectral image, or histogram, mode, in which spectral images were built up on board the telescope and downloaded as an integrated image.  The individual photon events were not preserved. Targets observed in this mode are labeled as ``HIST" observations in Tables~\ref{tab_list}, while time-tag observations are labeled ``TTAG."

We retrieved all of the \textit{FUSE} observations of CVs from the Multimission Archive (MAST) at Space Telescope Science Institute.  Spectra acquired before 2003 January were reduced using CalFUSE V2.4, the \textit{FUSE} data reduction pipeline.  All spectra acquired after that date were reduced using CalFUSE V3.2.  (Both versions of CalFUSE produce consistent spectral outputs, except in the case of background subtraction of faint targets; see below.) A detailed description of CalFUSE is given by \citet{dixon2007}. For each dataset, we inspected the data products (images of the count rate and aperture placement for each exposure and extracted spectra for each observation) to identify any problems with the observation and to ensure that the pipeline correctly screened out bad time intervals. For time-tag data, we combined the individual exposures in each segment using the CalFUSE tool \verb+ttag_combine+.  We then combined the eight time-averaged segment spectra into a single broadband spectrum using the IDL-based tool \verb+fuse_analysis+, provided on the \textit{FUSE} website.  Using the overlapping wavelength regions, we cross-correlated the spectra to correct for small wavelength offsets between channels and then combined the spectra, weighting by the error bars in each channel.  We also screened out bad spectral regions as needed (see below). The time-averaged spectra were binned to 0.1~\AA.  Finally, we combined spectra from different observations if they were acquired close in time and if there were no gross changes in the spectrum between the observations.

When examining the CalFUSE data products, we were searching for the effects of several consistent issues that appeared during \textit{FUSE} observations. For more information, see \cite{sahnow2000} and the \textit{FUSE} data handbook.  Here we discuss how we handled these effects for this survey:

\begin{enumerate}
\item{Jitter and channel drift:} Until 2005, telescope guiding was performed using a fine error sensor that imaged reflected light from the front surface of the LiF1 aperture block.  After July 2005, the prime guide channel was switched to LiF2. While objects generally remained well-centered in  the aperture of the optical channel used for guiding, the objects could drift out of  apertures of the other three channels, causing partial or total loss of target throughput. Both long term drifts (especially early in the mission) and more rapid jitter took place.  The longer term drifts were primarily due to changes in the thermal environment while the jitter was primarily due to true pointing jitter. For the latter, engineering information was used by CalFUSE to remove periods when the instrument was not in the aperture. In general, the jitter screening was accurate, although in inspecting the count rate plots, we occasionally found instances where the jitter screening intervals were too narrow compared to the actual drop in count rates. We corrected for this and for cases of drift where the target was still partially in the aperture by scaling the output spectra to the flux of the spectrum in the guide channel.  In some cases, an exposure was so heavily screened for drift that we would remove the exposure from the time-averaged spectrum entirely rather than keep a few tens of seconds of marginal data.  
\item{Event Bursts:} \textit{FUSE} regularly detected ``event bursts," large count rate events that were transient but could affect parts or all of one or more detectors. The source of the bursts is unknown.  CalFUSE detects and screens out bursts in time-tag observations by removing periods when the detected counts varied by more than five standard deviations from the mean count rate. In general, we found the burst detection procedure in CalFUSE to be accurate, so we rarely had to perform additional screening.
\item{Background subtraction:}  For faint targets (continuum $f_{\lambda} \lesssim 5\times10^{-14}$ ergs cm$^{2}$ s$^{-1}$ \AA$^{-1}$), CalFUSE often failed to properly subtract the background. This was particularly true for the final version of CalFUSE, V3.2, which used a three-component background model (a uniform component, daytime scattered airglow, and nighttime scattered light) that was often severely underconstrained for fainter targets. 
The background subtraction was most inclined to fail on the less sensitive SiC channels or on the edges of the LiF channels where sensitivity dropped off (especially $<1000$~\AA). Because the background contained both uniform and wavelength-dependent components, we were unable to easily correct for poor background subtraction (by, for example, setting the flux below the Lyman limit to 0).  Instead, we screened out regions that were unsuccessfully corrected. We regularly screened out the short wavelength ends of the LiF channels and the SiC1A and SiC2B segments, where we already had better coverage in the LiF channels.  We also screened out the other SiC segments for the faintest targets, leaving only the LiF channel coverage.  
\item{Detector Voltages:}  Occasionally no data were recorded on one or more detectors because they had been set to SAA, or low, voltage after single event upsets. We have noted such cases in the Appendices.
\item{The Worm:}  In order to improve quantum efficiency, sets of grid wires were placed just above the microchannel plates used in the detectors for \textit{FUSE}.  However, for certain positions of the target spectrum on the detector, the QE grid wires can shadow parts of the detectors, adversely affecting the flux calibration in these regions.  The most prominent feature, or ``worm," was located on the LiF1B segment.  The problem was not easy to calibrate out because the amount of shadowing depended on exactly where the spectrum was on the detector (see, for example, Fig.\ 11 of Dixon et al.). In most cases, we screened out the affected part of the LiF1B spectrum by comparing it to the LiF2A spectrum at the same wavelengths.  In a few cases, we could not do this because of lack of data in LiF2 due to channel drift or low detector voltages.  In those cases, we left the LiF1B spectrum uncorrected.
\end{enumerate}

The lists of CVs observed using \textit{FUSE} is given in Table~\ref{tab_list}. The table lists the object name, CV subtype, and orbital period and inclination (where known) for each target. The CV subtypes are given using the standard nomenclature as outlined in catalogs such as that of \citet{ritter2003}. Information about the systems was taken first from the catalog of Ritter \& Kolb \citet{ritter2003}\footnote{\url{http://physics.open.ac.uk/RKcat/}} and second (if needed) from the catalog of Downes \citep{downes2001}\footnote{\url{http://archive.stsci.edu/prepds/cvcat/index.html}}.  Note, however, that the classifications of individual systems can change in light of new observational data; these changes may not propagate to the general catalogs immediately. For each observation, we provide the data ID, start date, final exposure time (after CalFUSE screening), a flag indicating whether the observation was taken in time-tag or histogram mode, and a data quality estimate. Finally, we provide comments for certain observations (such as those observed in the MDRS channel) and a list of published references for each target. 
We have made the combined, time-averaged spectra available on MAST as a High Level Science Product (HLSP): \textit{FUSE} Survey of Cataclysmic Variables.  All of the spectra and plots of the data can be downloaded from the MAST HLSP archive\footnote{\url{http://archive.stsci.edu/prepds/fuse\_cv/}}.  The HLSP also includes details about the data calibration process for each target, particularly flagging cases where observational or calibration issues as discussed above affect the final data product.

\section{Analysis and Discussion} \label{sec_analysis}

\begin{figure}
\plotone{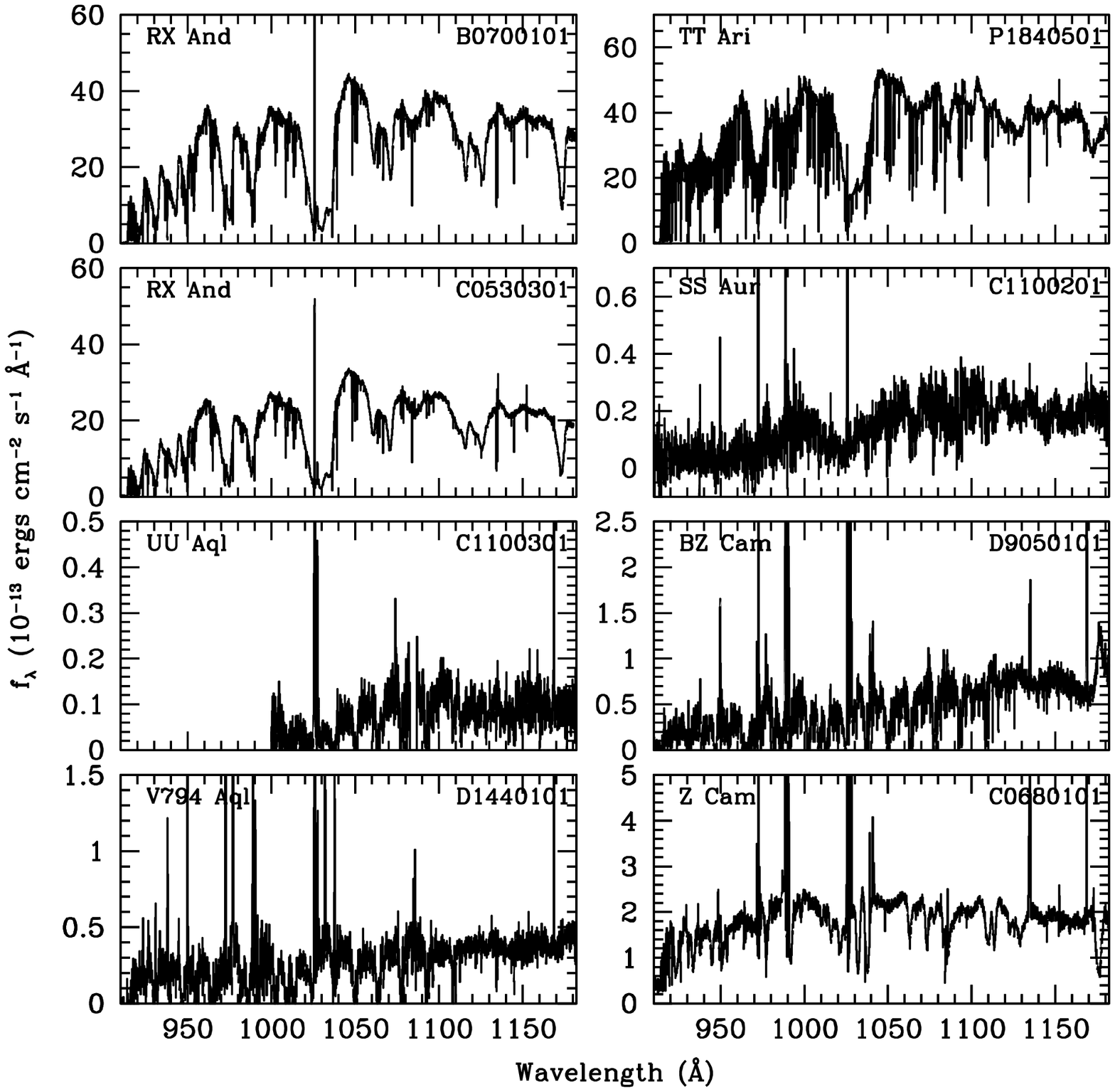}
\figcaption{The non-magnetic CVs observed by \textit{FUSE} shown in lexographical order.  The name of each target and the observation number are labeled.  Low signal to noise spectra are not shown and we show only one observed spectrum of each target unless the spectrum changed between observations. The narrow emission lines (as at Ly$\beta$ 1025) are due to terrestrial airglow. The plots are available in the electronic edition of the manuscript only.\label{fig_nonmags}}
\end{figure}

\begin{figure}
\plotone{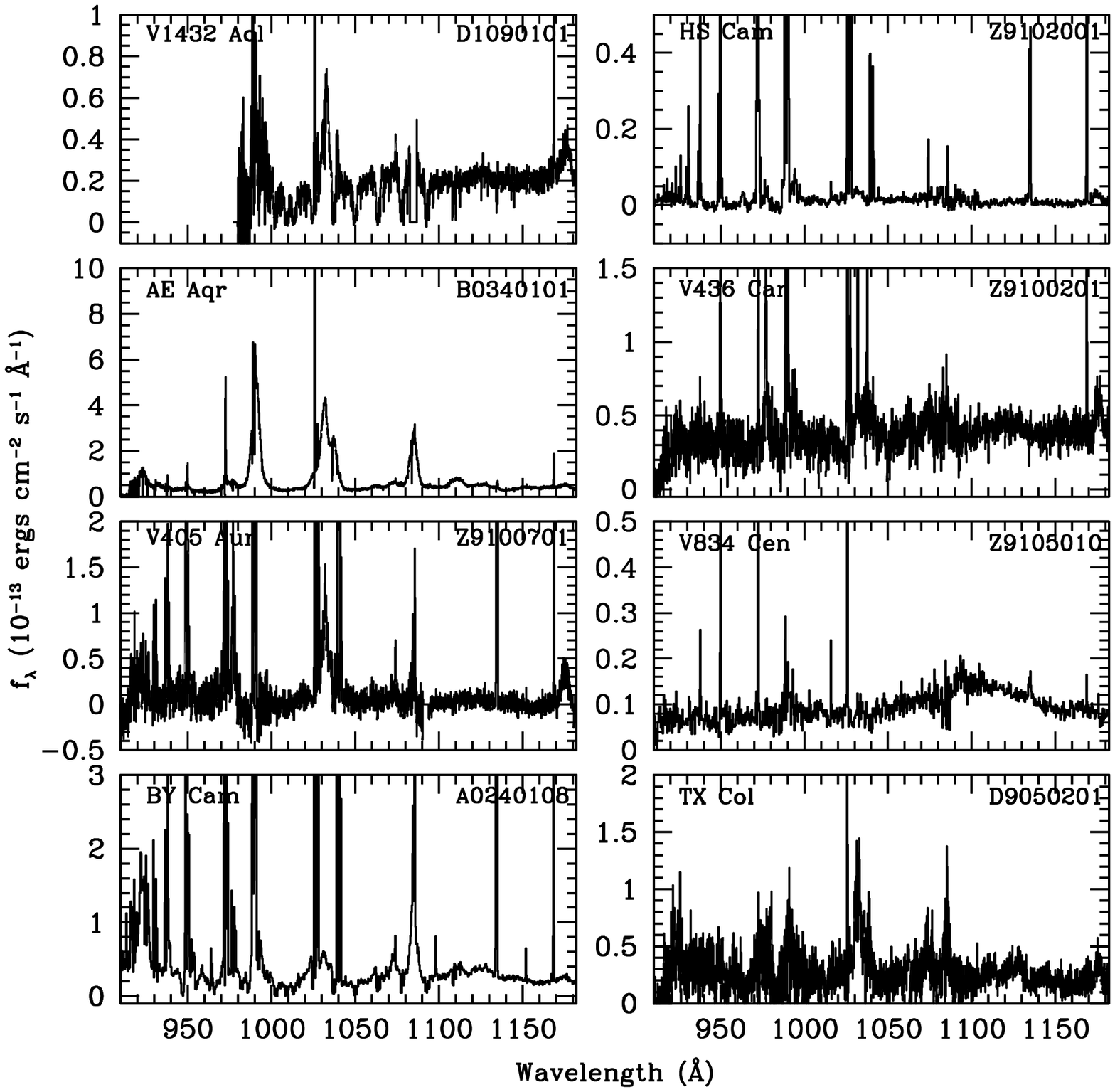}
\figcaption{Magnetic CVs observed by \textit{FUSE} shown in lexographical order. The name of each target and the observation number are labeled. Low signal to noise spectra are not shown and we show only one observed spectrum of each target unless the spectrum changed between observations. The plots are available in the electronic edition of the manuscript only. \label{fig_mags}}
\end{figure}

Figures~\ref{fig_nonmags} and~\ref{fig_mags} show the non-magnetic and magnetic CVs, respectively, observed by \textit{FUSE}. The spectra are given in lexicographical order.  We do not show an object spectrum if the signal to noise was low and we only show one spectrum per object unless the spectra changed dramatically between observations. The remaining spectra in the survey can be viewed on the MAST HLSP pages. 

Perhaps the key unifying characteristic of the FUV spectra of the CVs observed using \textit{FUSE} is their complexity.  Virtually every component of the system, with the exception of the cool donor star, is expected to emit in the FUV.  As a result, the FUV spectra are often superpositions of accretion disk, WD, and disk winds, with continuum and lines from multiple components. The FUV bandpass is rich in spectral features of a variety of species and ionization levels.  Many of these features, particularly of ionized species, appear in the spectra of CVs in both absorption and in emission (often in the same system). Typically, the lines are broad, with FWHM of hundreds of km~s$^{-1}$. In Table~\ref{tab_lines} we identify spectral features intrinsic to the CVs that have been seen in  \textit{FUSE} spectra. 

\begin{figure}
\plotone{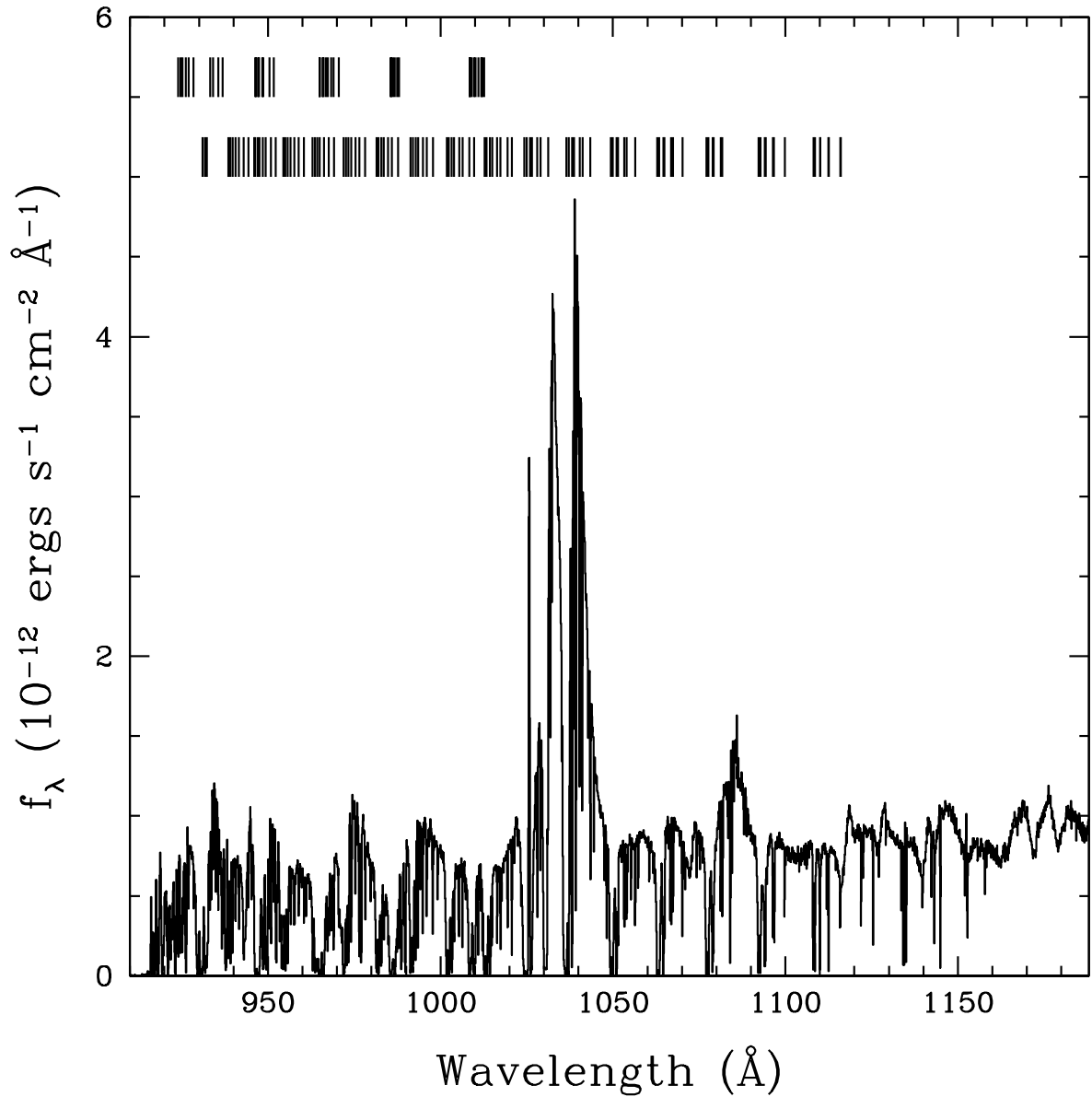}
\figcaption{The time-averged spectrum of V~Sge, showing the strong interstellar molecular hydrogen absorption typically seen in \textit{FUSE} spectra but uncommon in the spectra of nearby CVs. The upper and lower tick marks indicate the transitions of the Lyman and Werner H$_{2}$ rovibrational bands, respectively.\label{fig_vsge}}
\end{figure}

The spectra also show narrow absorption lines from neutral species originating in the interstellar medium along the lines of sight to the targets. In Table~\ref{tab_ism} we list the interstellar atomic transitions commonly seen in the \textit{FUSE} spectra of CVs.  Observable CVs are nearby objects, and therefore most do not show strong absorption from interstellar H$_2$, which appears in the \textit{FUSE} band and it nearly ubiquitous in \textit{FUSE} spectra of other targets.  Typically, when molecular H does appear in CVs, one sees only weak, unsaturated lines.  However, there are exceptions, such as the spectrum of V~Sge (Figure~\ref{fig_vsge}), which shows  strong, saturated H$_{2}$ absorption.  Finally, most spectra show emission lines from terrestrial airglow.  Detailed information about FUV airglow emission can be found in \citet{feldman2001}.

In the following sections, we discuss the characteristics of the \textit{FUSE} CV survey as a function of CV subtype.

\subsection{Dwarf Novae in Quiescence}

Thirty-eight DN were observed using \textit{FUSE}. Thirty-seven of the systems were observed in quiescence.  One system (RX~And) was caught only in its outburst state.  For some of the systems, particularly those observed in extended \textit{FUSE} campaigns (e.g., SS~Cyg, U~Gem, VW~Hyi, and WZ~Sge), we know when the DN were in quiescence from ground-based monitoring carried out by organizations such as the AAVSO. For the other systems, we classified the observations as being in quiescence based on their fluxes and spectral shapes: e.g., relatively faint fluxes, flat or red continua, emission spectra and/or absorption lines characteristic of a WD.

\begin{figure}
\plotone{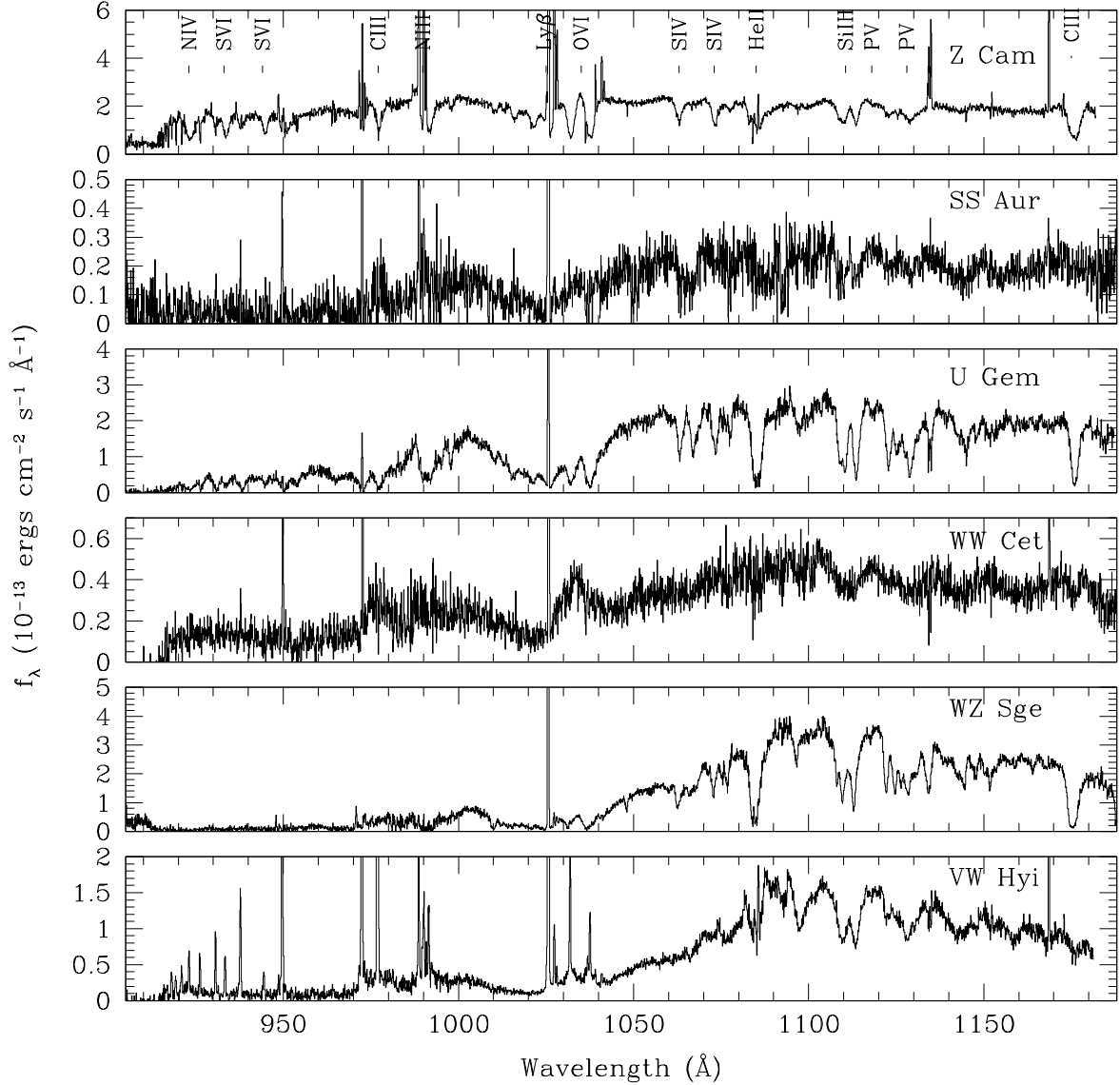}
\figcaption{Quiescent dwarf novae with spectra dominated by the white dwarf. Shown from top to bottom are Z~Cam (T$_{WD}$ = 57,000~K), SS~Aur (T$_{WD}$ = 33,000~K), U~Gem (T$_{WD}$ = 30,000~K), WW~Cet (T$_{WD}$ = 26,000~K), WZ~Sge  (T$_{WD}$ = 23,000~K), and VW~Hyi (T$_{WD}$ = 20,000~K). The WD temperatures are the best-fit single-component models from \citet{hartley2005}, \citet{sion2004a}, \citet{long2006}, \citet{godon2006}, \citet{long2003}, and \citet{long2009}, respectively \label{fig_dnwd}}
\end{figure}

\begin{figure}
\plotone{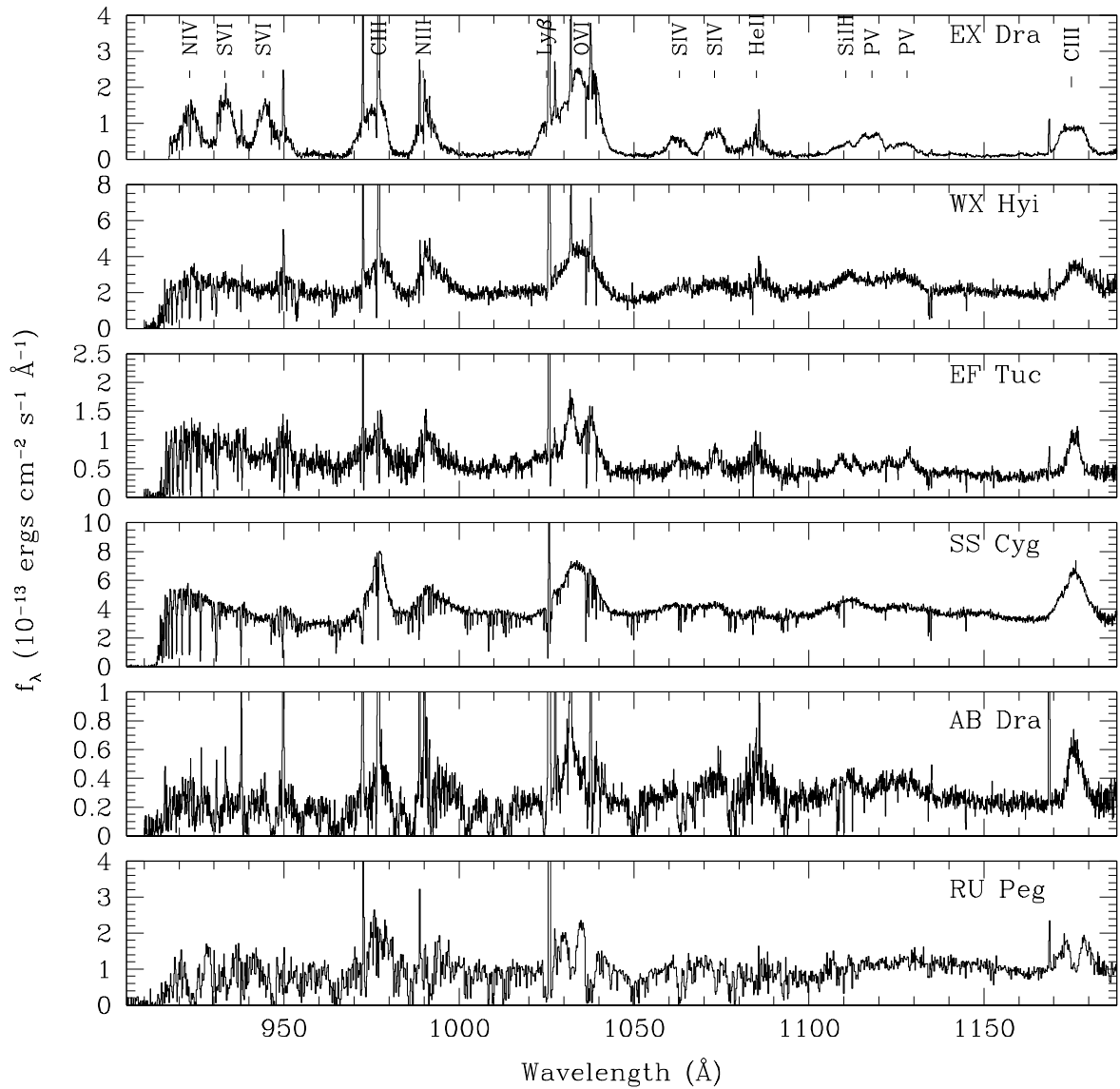}
\figcaption{Quiescent dwarf novae with strong emission line spectra. Shown from top to bottom are the spectra of EX~Dra (D9050501), WX~Hyi (U1062001), EF~Tuc (E9890701), SS~Cyg (P2420101), AB~Dra (D9050301), and RU~Peg (C1100101). \label{fig_dnem}}
\end{figure}

The quiescent DN spectra can be broadly divided into two categories: those that show broad absorption in the Lyman lines and narrower absorption in some metal lines, and those that have flat continua and emission lines.  The former are those systems in which the WD dominates the FUV spectrum while in the latter, emission from the accretion disk obscures the WD signatures. Figures~\ref{fig_dnwd} and~\ref{fig_dnem} show examples of the two categories. 
There is overlap between the two cases, of course, including emission lines in ``WD spectra''  (such as the \ion{O}{6} emission in WW~Cet) and ``emission line systems'' with signatures that suggest an underlying WD component (such as the absorption core in \ion{C}{3} $\lambda$1175 in RU~Peg).

Figure~\ref{fig_dnwd} shows the spectra for several WD-dominated systems in order of decreasing WD temperature. The WD temperatures are from single-component WD model fits to the spectra. WZ~Sge had just entered quiescence after a superoutburst when the \textit{FUSE} spectrum was taken and the WD model fit gives a temperature of T$_{WD}$ = 23,000~K; the temperature late in quiescence is lower, T$_{WD} \simeq$16,000~K \citep{long2003,godon2004a}.  There have been numerous papers devoted to fitting WD models to \textit{FUSE} spectra of quiescent DN (see the citations in Table~\ref{tab_list}). The aims of these studies have been to obtain WD parameters (temperatures, masses, gravities), decompose the multiple emission sources in the systems (including multiple components on the WD surface), trace the effects of active accretion on the structure of the boundary layer and the WD atmosphere, and probe how the WD responds to enhanced accretion after outbursts. 

One consistent conclusion of these projects is that the FUV spectra are often better fit by multi-component models than single temperature WD models. In some cases, the spectra are fit with two-temperature WD models, interpreted as the bulk of the WD having one temperature while some fraction of the WD surface is hotter \citep[e.g.,][]{froning2001,hartley2005,godon2006,long2006,sion2007,long2009}. 
One explanation for this picture is that a hot ``accretion belt'' is spun up on the WD surface by the accretion of disk material \citep{kippenhahn1978,sion1996,cheng1997,long2003}.  However, none of the data have shown unambiguous evidence of rapid rotation uniquely consistent with a belt, leaving the source of the second component unconstrained \citep{long2006}. Other analyses have used WD plus steady-state accretion disk models \citep[e.g.,][]{godon2009,sion2010}, although generally the inclusion of the disk component does not improve the fit (see below for more discussion on this point). \citet{long2009} followed VW~Hyi through decline from a superoutburst to the next normal outburst. They found that the FUV spectrum is dominated by the WD, which cools during the interoutburst period. They also found that two additional components were necessary to model the spectrum: a variable component modulated on the orbital period, modeled by a thermal spectrum, and attributed to the hot spot; and a third component that appears long after outburst and is composed of emission lines plus a pseudocontinuum, which they attributed to the disk.  \citet{godon2005}, on the other hand, pointed to the boundary layer as the source of the additional emission in VW~Hyi. 

A number of systems, including U~Gem, VW~Hyi, EY~Cyg, CH~UMa, and SS~Aur, show abundance anomalies indicative of CNO processing in the accreted material. The source of the abundance differences from solar abundance ratios are not known but have been attributed either to nuclear evolution of the donor star or to pollution of the donor by previous nova explosions \citep{marks1998,sion2001,schenker2002}. To date, the only definitive abundance variations are decreased (relative to solar) carbon and enhanced nitrogen in the FUV spectra, but there are hints that other elements may show abundance anomalies  \citep[e.g., low Si in SS~Aur;][]{sion2004a} that will help distinguish between the models.

Figure~\ref{fig_dnem} shows examples of emission line-dominated quiescent DN.  EX~Dra has a unique spectrum consisting of strong line emission and little to no continuum. Its binary inclination ($i = 84.2^{\circ}$) is such that the system is nearly edge-on and is therefore dominated by the emission line region, typically ascribed to a photoionized accretion disk chromosphere (though this region and indeed the vertical structure of low-state disks in general is poorly understood). The other quiescent DN with emission line spectra are more similar to each other. They exhibit strong, broad emission lines, particularly in \ion{C}{3}, \ion{N}{3}, and \ion{O}{6}, and flat to blue continua. The sources of the continuum emission are difficult to characterize.  \citet{sion2004a}, for example, fit the spectrum of RU~Peg with a 53,000~K WD, but the model did not fit the spectrum well, particularly in the blue. They were unable to improve the fit by switching to a steady-state accretion disk model or a hybrid WD/disk model. The lack of robust descriptions of the structure and properties of quiescent accretion disks hinders modeling efforts: eclipse maps of eclipsing quiescent DN show that their disks are not in a steady state \citep{horne1993}, so it is unsurprising that steady-state disk models do not provide good descriptions of the FUV continuum spectrum. As a result, successful models of the FUV spectra of  emission line dominated systems remain elusive. (For a more detailed discussion on this point, see Long et al.\ 2005.)

\subsection{Dwarf Novae in Outburst}

\begin{figure}
\plotone{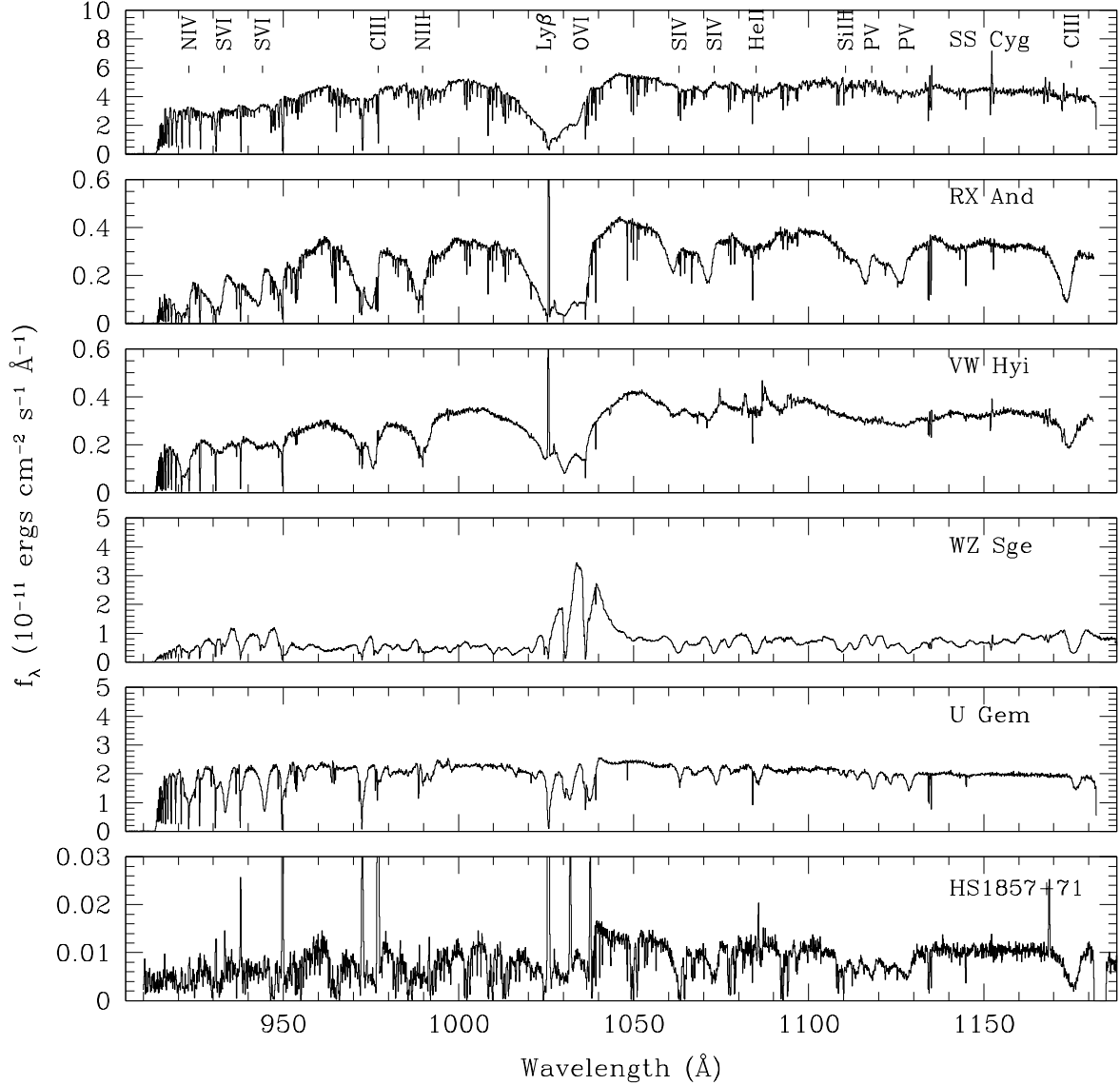}
\figcaption{The time-averaged spectra of six dwarf novae observed in outburst.  Shown from top to bottom are the spectra of SS~Cyg ($i = 37^{\circ}$), RX~And ($i = 51^{\circ}$), VW~Hyi ($i = 60^{\circ}$), WZ~Sge ($i = 75^{\circ}$),  U~Gem ($i = 70^{\circ}$), and HS1857+7127 (inclination unknown). The prominent spectral features are labeled in the top panel.  The spectra are taken from observations A1260270 (SS~Cyg), B0700101 (RX~And), E1140111 (VW~Hyi), Z0030101 (WZ~Sge), A1260112 (U~Gem), and U1041103 (HS1857+7127). \label{fig_dno}}
\end{figure}

RX~And, SS~Cyg, AB~Dra, U~Gem, VW~Hyi, WZ~Sge, and HS1857+7127 were observed in outburst.  Their outburst states were confirmed by examining their AAVSO light curves for all targets but the last. The AAVSO observations of HS1857+7127 are too sparse to confirm its optical state, but the 2007 Mar 09  \textit{FUSE} spectrum was $\simeq$10 brighter than the previous one taken on 2006 Sep 13 and is probably an outburst observation. For SS~Cyg, U~Gem, VW~Hyi, and WZ~Sge, \textit{FUSE} obtained extended campaigns tracking the CVs through outburst and decline.  The VW~Hyi campaign observed the system both in superoutburst and in the subsequent normal outburst. Fig~\ref{fig_dno} shows the outburst spectra for six systems.   The U~Gem, VW~Hyi, and WZ~Sge observations from outburst to quiescence have been analyzed individually in previous papers \citep{long2009,froning2001,long2003}.

Taken as a set, the intrinsic FUV spectra of the DN in outburst show both morphological similarities and variations.  SS~Cyg, RX~And, and VW~Hyi are similar, with spectra characterized by fairly flat continua longward of 1000~\AA, declining fluxes in the blue, and strong, broad (FWHM$\simeq$1000--5000~km~s$^{-1}$) absorption lines of \ion{H}{1} and metals.  The U~Gem and HS1857+7127 spectra have flat continua over the full \textit{FUSE} range punctuated by relatively narrow (FWHM$ \sim$ 500~km~s$^{-1}$) absorption lines in \ion{H}{1} and ionized metals. The AB~Dra spectrum is noisy (it was only observed for 600 sec and has interstellar molecular H$_{2}$ contamination) but the \ion{O}{6} and \ion{C}{3} $\lambda$1175 lines are  broad and blueshifted. Finally, the WZ~Sge spectrum has a flat continuum with a combination of absorption and emission components, punctuated by very strong, broad emission in \ion{O}{6}.  

Many of the morphological variations can be ascribed to viewing inclination, which increases from more face-on to increasingly edge-on from top to bottom in Figure~\ref{fig_dno}. In particular, if the spectral line features are ascribed to a strong disk wind, then the transition from broad, blue-shifted absorption in SS~Cyg, RX~And, and VW~Hyi to the spectrum of WZ~Sge, which contains lines with P~Cygni profiles (\ion{S}{6}, for example) and the strong \ion{O}{6} doublet emission profile, is consistent with the picture of a bipolar outflow seen in absorption against the disk continuum at low inclination and in emission in scattered light at high inclination. \citet{froning2002} found a good fit to the spectrum of SS~Cyg with a model consisting of a steady-state accretion disk plus a rotating, biconical disk wind, wherein roughly 1\% of the accreted material escapes in the outflow.  However, the spectra of U~Gem and HS1857+7127 complicate the picture. Their absorption spectra consist of narrow lines at low velocities ($\leq 700$~km~s$^{-1}$) that vary on the orbital phase, which is inconsistent with an escaping wind. \citet{froning2001} attributed the absorption in U~Gem to vertically-extended material in the outer accretion disk, probably associated with a disk bulge or mass accretion stream overflow.

For SS~Cyg, U~Gem, and WZ~Sge, mass accretion rates in the disk at outburst peak have been determined using model fits of steady-state accretion disks to the FUV continuum, giving \.{m}$_{disk} = 4.4\times10^{-9}$~\msun~yr$^{-1}$, \.{m}$_{disk} = 6.9\times10^{-9}$~\msun~yr$^{-1}$, \.{m}$_{disk} = 8.5\times10^{-10}$~\msun~yr$^{-1}$, respectively \citep{froning2002,froning2001,long2003}. The outburst accretion rates of U Gem and SS Cyg, which both have periods above the period gap, are $\sim$5--8 times higher than that of the short-period WZ Sge. In all three cases, the model disk spectra provided good qualitative fits to the continuum shape, with the exception of the shortest wavelengths ($\lambda < 950$~\AA), which shows excess emission in the observed spectra compared to the models. The discrepancy could be caused by an additional emission source, such as the boundary layer or the hot spot where the accretion stream impacts the disk. The excess emission (over a white dwarf component) persists in quiescence in U~Gem and VW~Hyi; in the latter, the excess component is modulated on the orbital phase, which leads \citet{long2009} to attribute it to the hot spot. 

\subsection{Novalikes}

Thirty-two NLs were observed.  This total includes V~Sge, which is classified as a CV but may instead be a colliding wind binary or a supersoft X-ray binary \citep{patterson1998,wood2000}; and CQ~Dra, which is a triple star system that has been classified as both a symbiotic star and a CV, with its X-ray properties favouring the former hypothesis \citep{hric1991,wheatley2003}. The NL designation for CVs is broad, encompassing any system that does not show DN outbursts and does not have signatures of the presence of a magnetic WD.  As a result, it is not surprising that the FUV spectra of NLs are among the most diverse and complex spectra of CVs observed by \textit{FUSE}. Here, we will highlight some of the broad groupings that appear and what they reveal about the the FUV properties of NLs.

\begin{figure}
\plotone{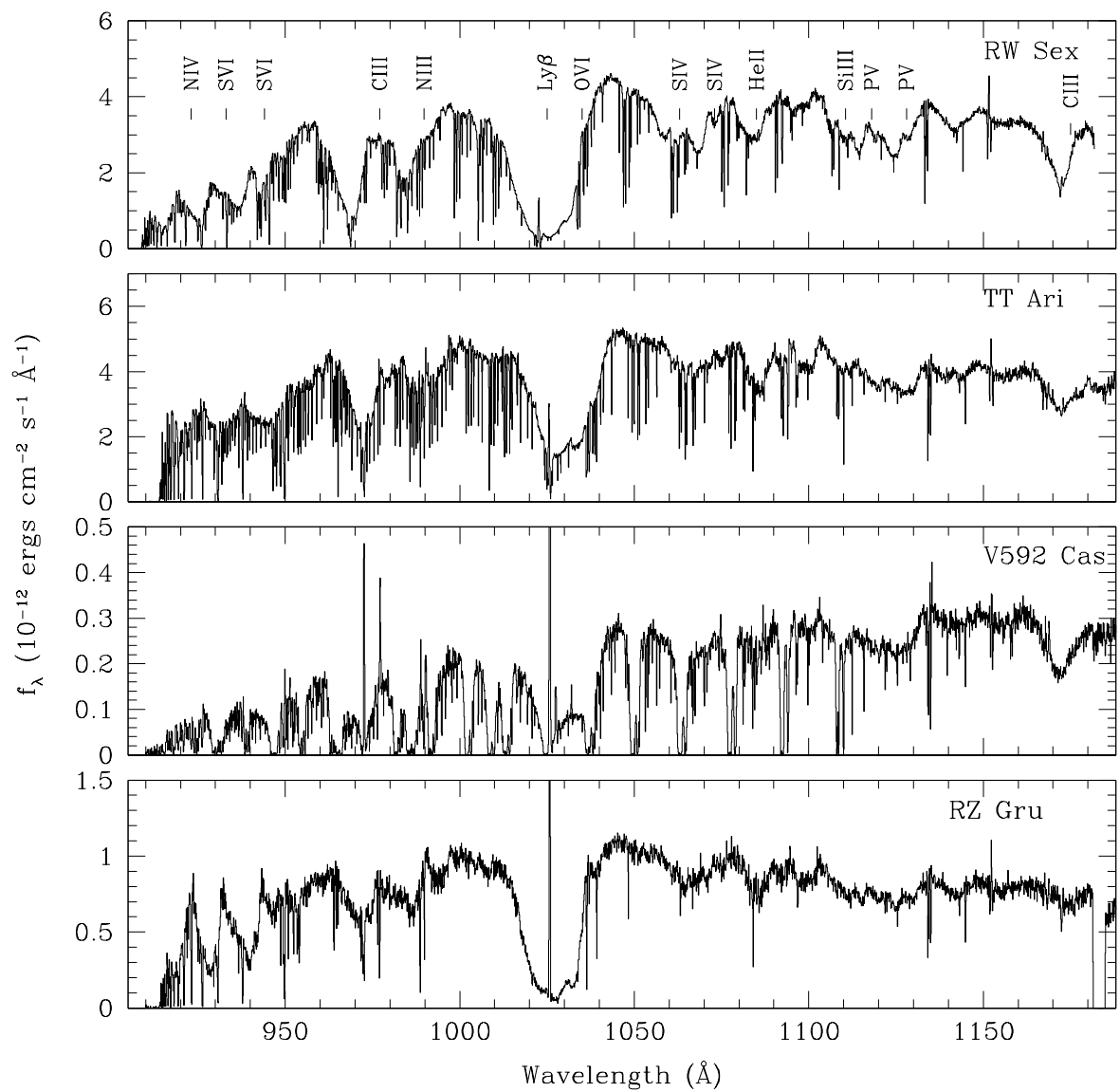}
\figcaption{Novalike CVs with strong, broad absorption line spectra.  Shown from top to bottom are the spectra of RW~Sex, TT~Ari, V592~Cas (observation D1140103), and RZ~Gru. \label{fig_nlabs}}\
\end{figure}

Figure~\ref{fig_nlabs} shows several NLs that share similar absorption spectra. The spectrum of V592~Cas is heavily contaminated by interstellar H$_{2}$ but its \ion{C}{3} $\lambda$1175 line has a FWHM of  2000~km~s$^{-1}$, which suggests that it is part of this group.  All the targets are low inclination, $10^{\circ} \leq i \leq 34^{\circ}$.  The spectra of these four targets are characterized by strong, blueshifted absorption lines of \ion{H}{1}, \ion{He}{2}, and ionized metals. In RZ~Gru, the absorption feature at 1025~\AA\ has a FWHM of 5200~km~s$^{-1}$, although the feature is likely a blend of \ion{H}{1} and the \ion{O}{6} doublet.  The \ion{S}{6} lines are more isolated and easier to measure unambiguously: they have FWHM$\simeq$2000~km~s$^{-1}$ and are blueshifted at line center by 1700~km~s$^{-1}$.  Similarly, the FWHM of the lines in TT~Ari are $\simeq$2200~km~s$^{-1}$ \citep{hutchings2007}.  In RW~Sex and V592~Cas, the lines are highly variable. In RW~Sex, the variability is modulated on the orbital period, with all of the lines shifting from deep lines with maximum absorption at $-1000$~km~s$^{-1}$ to much weaker lines centered on zero velocity \citep{prinja2003}.  The variability is confined to velocities blueward of the rest velocities of the lines. Similarly, the line variability in V592~Cas is strictly modulated on the orbital period and not on the superhump periods in the system \citep{prinja2004}. 

The spectra of these targets have been interpreted as originating in strong winds outflowing from the accretion disk. The underlying continuum is dominated by emission from the disk. \citet{linnell2010} modeled the continuum in RW~Sex with steady state disk models with  \.{m}$_{disk} \simeq 6\times10^{-9}$~\msun~yr$^{-1}$, though they found that the disk plus WD models could not fit the spectral shape over the full UV range (\textit{FUSE} plus HST and IUE data).  Their preferred model utilized a more shallow temperature gradient in the disk than that given by the standard disk model,  $ T\propto R^{-\frac{3}{4}}$ \citep{pringle1972}.  The absorption line variability in RW~Sex and V592~Cas indicates asymmetries in the outflows. Interestingly, the continuum does not vary in these targets, indicating that changes in the wind structure are not tied to the underlying disk. The lack of line variability modulated on the superhump periods in V592~Cas points to the same conclusion. The source of the asymmetry in the outflow is unknown, though Prinja et al. speculate that the wind may be rotating obliquely relative to the observer's line of sight.

\begin{figure}
\plotone{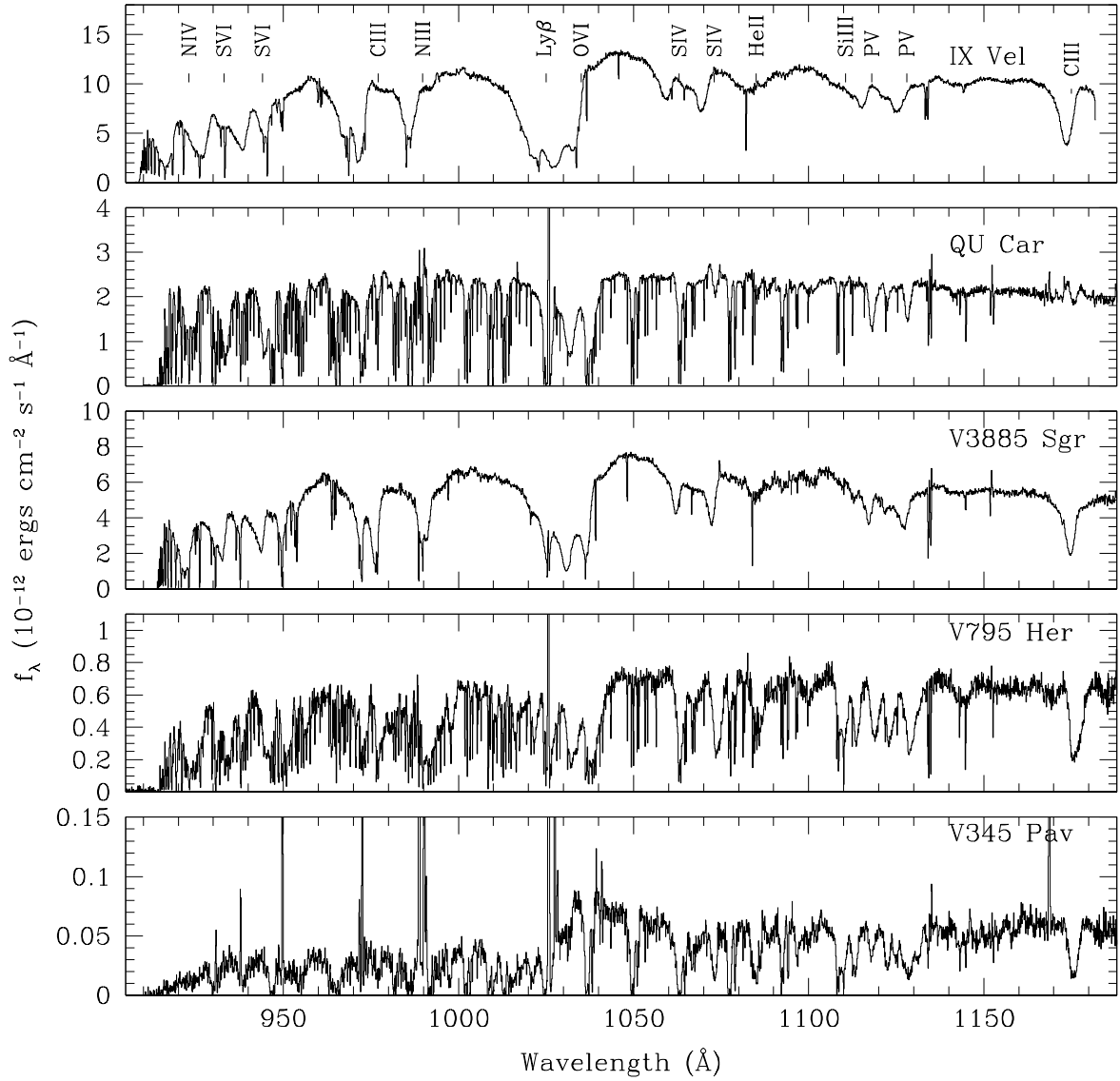}
\figcaption{Novalike CVs with relatively narrow absorption line spectra.  Shown from top to bottom are the spectra of IX Vel (Q1120101), QU~Car (D1560101/02), V3885~Sgr (D9051501/02), V795~Her, and V345~Pav (D9131101). \label{fig_nlnar}}
\end{figure}

Many of the NL FUV spectra have absorption lines that are narrower and with lower blueshifts than those discussed above. Figure~\ref{fig_nlnar} shows five examples.  Similar systems include BB~Dor, BP~Lyn, MV~Lyr (in its high state), AH~Men, and FY~Per.  The binary inclinations are not known for most of these sources although they generally may be higher than those discussed above: e.g., IX~Vel has $i = 60^{\circ}$  and QU~Car has $i < 60^{\circ}$. V3885~Sgr has been alternately described as $i < 50^{\circ}$ and $i = 60^{\circ}-70^{\circ}$, though \citet{hartley2002} prefer the latter because of the similarity of its UV spectrum to that of IX~Vel. However, MV~Lyr is at low inclination, $i = 12^{\circ}$, so the morphological changes in the spectra are probably also tied to physical variations in the outflows. BB~Dor may also have a low inclination \citep{godon2008}. In the two systems for which steady state disk models have been fit to the \textit{FUSE} spectra, the mass accretion rates are comparable to those seen in the other NLs: \.{m}$_{disk} \simeq10^{-9}$~\msun~yr$^{-1}$ for BB~Dor and \.{m}$_{disk} = 5\times10^{-9}$~\msun~yr$^{-1}$ for V3885~Sgr \citep{godon2008,linnell2009}. 

\citet{prinja2003} compared the UV spectra of several NLs. They note that systems like IX~Vel and V3885~Sgr not only have narrow lines with absorption minima at lower velocities ($-500$~km~s$^{-1}$ vs.\ $-1000$ to $-2000$~km~s$^{-1}$ for \ion{C}{4} $\lambda$1550) than systems like RW~Sex, they also show less variability in the lines. In the \textit{FUSE} spectra, the systems shown in Figure~\ref{fig_nlnar} have line widths of FWHM$\simeq$500--1500~km~s$^{-1}$ and are centered at the rest velocity of the line or only slightly blueshifted by $<300$~km~s$^{-1}$.  \citet{godon2008} fit the spectrum of BB~Dor with composite WD+disk models. Although the model spectra (dominated by the disk component) provided good fits to broader absorption features like the broad hump centered on Ly$\alpha$ and \ion{C}{3} $\lambda$1176, the other spectral lines are too narrow to be consistent with the disk or WD models. Instead, the lines in these NLs resemble those seen in U~Gem in outburst \citep{froning2001} and are likely associated with vertically extended material above the disk, although whether this material is flowing out of the system as in a more traditional wind or whether it is more consistent with a disk chromospheric region is not clear.

\textit{FUSE} observed two NLs from the AM~CVn subclass: V803~Cen and AM~CVn itself. The AM~CVn stars are believed to be ultra-compact systems in which both the donor star and the accretor are degenerate objects (see discussion in Warner 1995). AM~CVn persists in a stable high state, while V803~Cen is occasionally observed in a low state.  The FUV spectra of both targets are quite similar, suggesting that V803~Cen was in outburst when it was observed with \textit{FUSE}. The spectra show flat continuum emission and blueshifted absorption lines of \ion{He}{2} and metals with FWHM$\sim$1500~km~s$^{-1}$. The \ion{O}{6} line in V803~Cen may have a P~Cygni profile but the spectrum is quite complex so this cannot be stated definitively.  Despite their comparable continuum shapes, AM~CVn has stronger high ionization transitions, like \ion{S}{6}, while V803~Cen has stronger absorption in lower ionization species such as \ion{C}{3} and \ion{Si}{3}.  This suggests that the ionization structure of the wind may be more affected by the WD in the more compact AM~CVn system.

\begin{figure}
\plotone{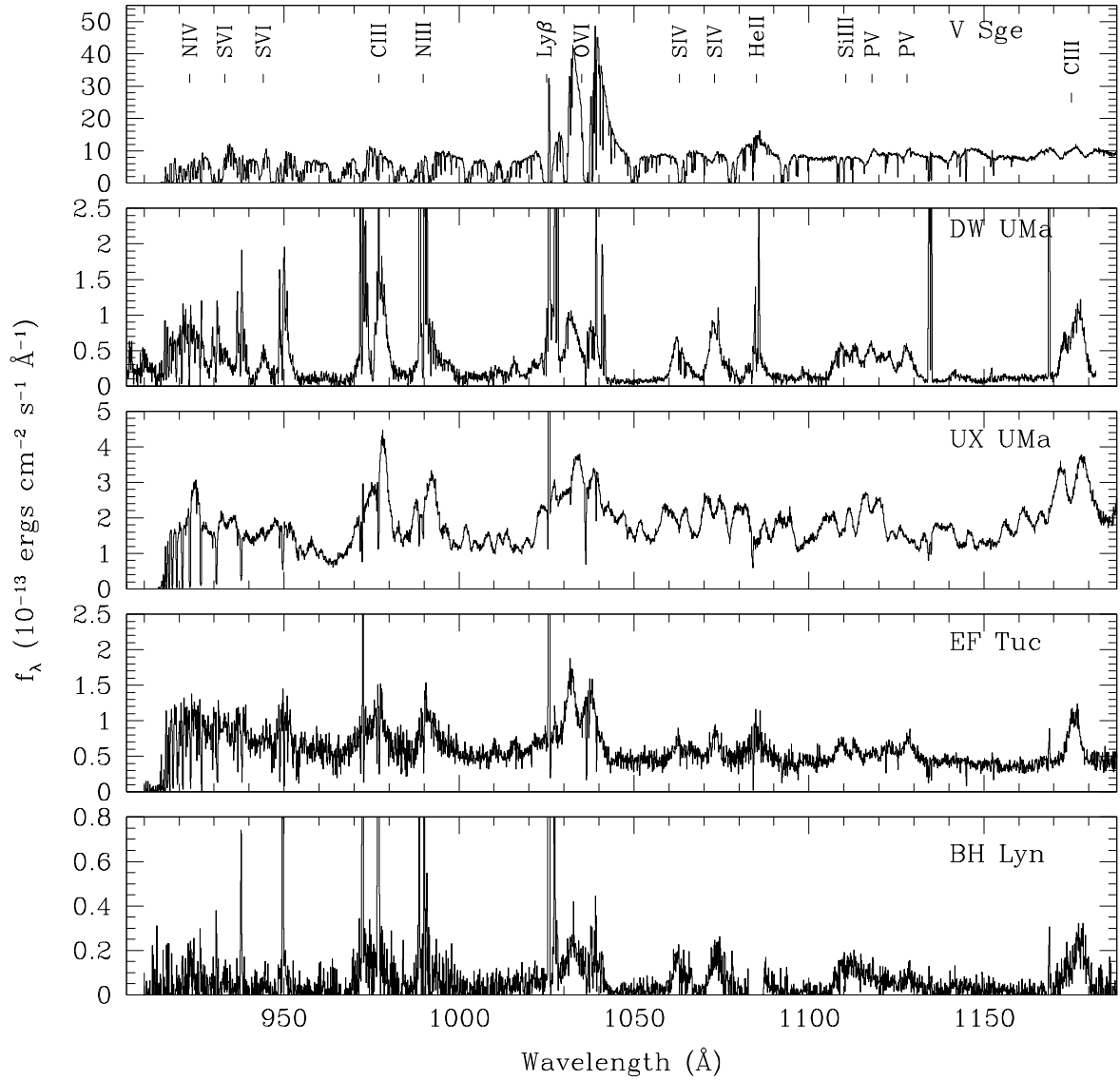}
\figcaption{Novalike CVs with emission line spectra.  Shown from top to bottom are the spectra of V~Sge (B0430101), DW~UMa, UX~UMa (B0820101/02), EF~Tuc (E9890701), and BH~Lyn. \label{fig_nlem}}
\end{figure}

Figure~\ref{fig_nlem} shows NL spectra with lines in emission. Other systems that show emission lines are BZ~Cam, CQ~Dra (though see below), RXJ0625+7334 (which is however likely an IP; Araujo-Betancor et al.\ 2003), SDSS080908+3841 and 0922+4950. Many of these are high inclination systems: V~Sge ($i=90^{\circ}$), DW~UMa  ($i=80^{\circ}$), and UX~UMa ($i=71^{\circ}$). BZ~Cam is low inclination ($i<40^{\circ}$) but the \ion{C}{3} $\lambda$1176 line (the rest are lost in interstellar H$_{2}$) has a P~Cygni profile.  Some spectra are purely in emission while some are mixtures of emission and absorption components. Using spectra taken in and out of eclipse, \citet{froning2003} decomposed the relative components of the spectrum of UX~UMa.  The eclipse spectrum, which is from vertically-extended, uneclipsed material (i.e., the disk wind), is an emission spectrum similar to that of DW~UMa. The eclipsed material is in absorption, which the authors attributed to an optically thick chromosphere close to the disk plane.  The  FUV spectrum of V~Sge is very similar to that of WZ~Sge at outburst peak, showing very strong \ion{O}{6} emission from an outflow viewed edge-on.  The spectra also can be highly variable, even out of eclipse: \citet{hoard2003} found variations in both lines and continuum modulated on the orbital phase, probably associated with overflow of the mass accretion stream after the initial impact point. In the high inclination systems, the continuum can be challenging to model.  In DW~UMa, for instance, a flared disk rim is believed to shield most of the inner accretion disk and WD from view \citep{knigge2000}, so standard disk models (which assume a thin, flat disk) do not fit the  FUV continuum well \citep{hoard2003}.  Both \citet{froning2003} and \citet{linnell2008} found that disk and disk plus WD spectra do not fit the FUV spectral shape in UX~UMa well either, even when (in the latter study) the longer wavelength UV-optical spectra are well fit. 

Finally, we briefly discuss some unusual NLs observed by \textit{FUSE}. a) MV~Lyr was observed twice with \textit{FUSE}. The second observation (D9050901) was a fairly standard NL spectrum, but the first observation (C0410301) caught the system in a low state, revealing the $T_{WD} = 47,000$~K WD \citep{hoard2004}. b) P831-57 has the spectrum of a WD, so it is either at VY~Scl NL caught in a low state or has been mis-classified as a NL. c) The spectrum of CQ~Dra resembles that of a WD with emission lines with narrow cores (FWHM$\sim$350~km~s$^{-1}$) and broader wings superimposed. Its spectrum is similar to the \textit{FUSE} spectra of symbiotic stars \citep[e.g.,][]{young2005,crowley2008}, consistent with the identification of CQ~Dra as a symbiotic star based on X-ray observations \citep{wheatley2003}. d) SDSS015543+0028 is classified as a NL but probably is actually a magnetic polar \citep{szkody2002,woudt2004}.

\subsection{Intermediate Polars}

\begin{figure}
\plotone{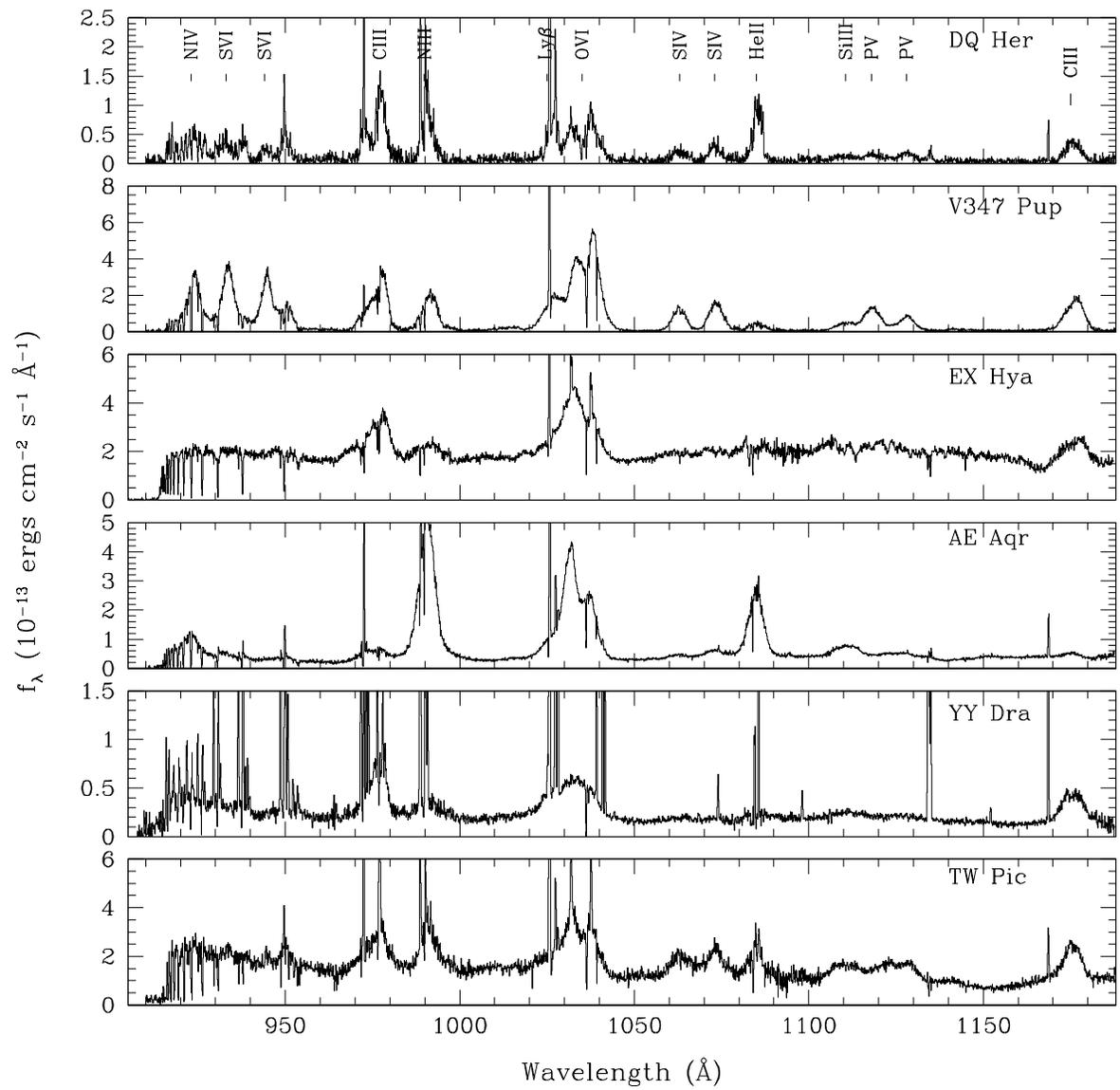}
\figcaption{The time-averaged spectra of six intermediate polars observed with \textit{FUSE}.  Shown from top to bottom are the spectra of DQ Her ($i = 86.5^{\circ}$), V347 Pup ($i = 84^{\circ}$), EX Hya ($i = 78^{\circ}$), AE Aqr ($i = 58^{\circ}$), YY Dra ($i = 45^{\circ}$), and TW~Pic. The prominent emission lines are labeled in the top panel.\label{fig_ip}}
\end{figure}

\textit{FUSE} observed twelve intermediate polars, designated ``NL IP" in Table~\ref{tab_list}.  In addition, RXJ0625+7334 is not classified as an IP in the general CV catalogs but is also in this class \citep{araujo2003}. All of the targets were detected, although only \ion{O}{6} and \ion{C}{3} emission can be seen in the spectrum of V709~Cas. Figure~\ref{fig_ip}  shows the spectra of six intermediate polars:  DQ~Her, V347~Pup, EX~Hya, AE~Aqr, YY~Dra, EX~Hya, and TW~Pic (but note Norton et al.\ 2002 who argue that TW Pic may not be an IP).  The spectra are organized from top to bottom in order of decreasing inclination. (The inclination of TW~Pic is unknown but may be low based on the lack of X-ray pulsations;  Norton et al. 2002)

\begin{figure}
\plotone{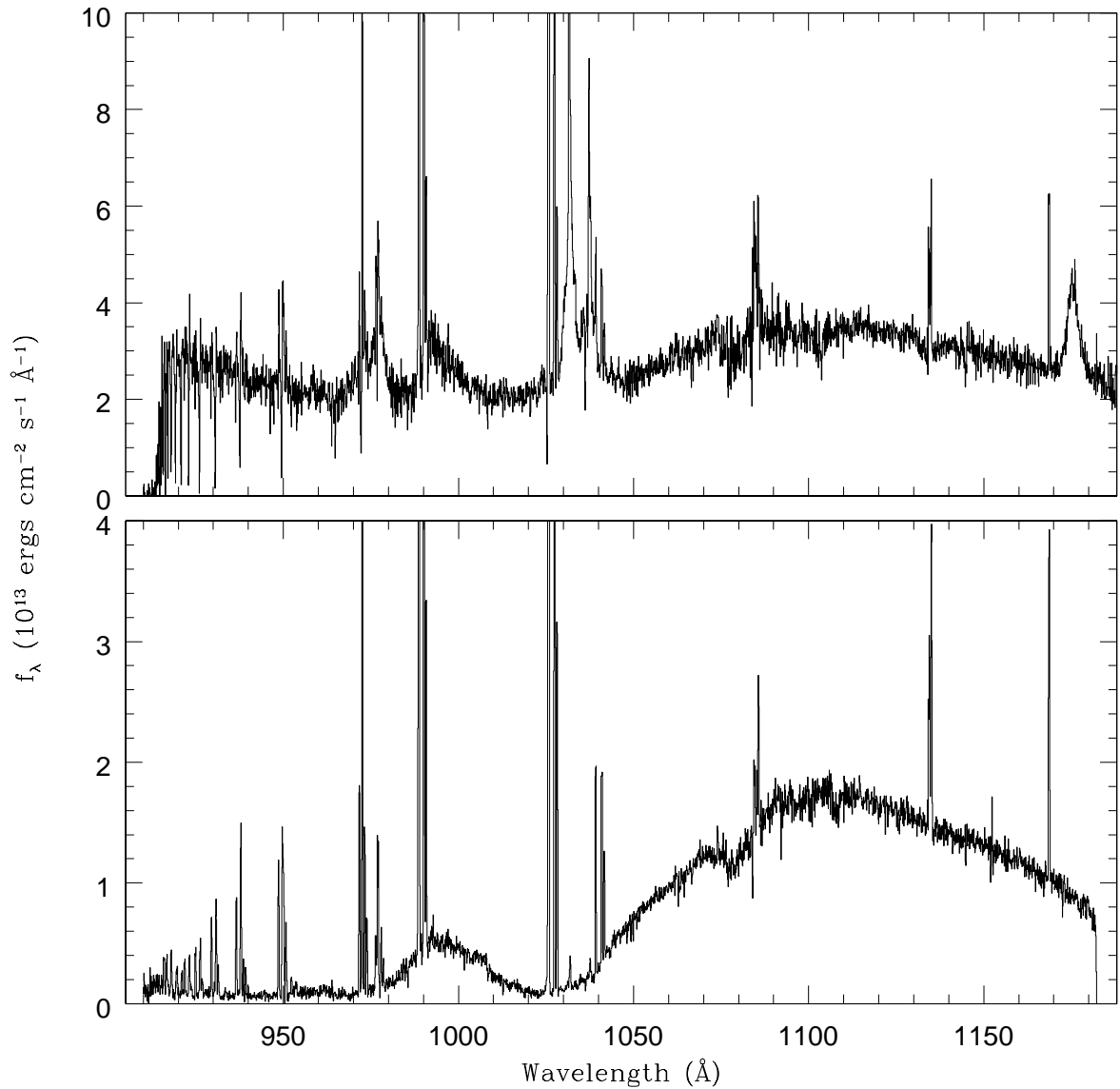}
\figcaption{The time-averaged spectrum of the polar AM~Her taken in two different epochs.  The top panel shows the Z0060101 observation of AM~Her, acquired in 2002 May 11.  The bottom panel shows the P1840601 observation of AM~Her, acquired in 2000 May 12, when the system was in a low accretion state.\label{fig_amher}}
\end{figure}

The FUV spectra of the IPs are similar.  They all (with one exception; see below) show strong line emission in \ion{C}{3}, \ion{N}{3}, and  \ion{O}{6}.  Several of the systems also have emission from \ion{H}{1}, \ion{He}{2}, \ion{Si}{3}, \ion{N}{4}, \ion{S}{4}, \ion{P}{5}, and \ion{S}{6}. The line strengths and EWs vary from system to system. This is only weakly linked to inclination.  The high inclination systems DQ~Her, V347~Pup, and AE~Aqr have very strong emission line spectra and weak continua but EX~Hya, at an inclination of $i=78^{\circ}$, has a strong continuum and relatively weak lines. EX~Hya is the only IP observed with an orbital period below the period gap. Its continuum spectrum is unusual, with a hump at 1100~\AA\ that is reminiscent of a cooler WD spectrum (e.g., the spectra of WW~Cet, WZ~Sge, and VW~Hyi in Figure~\ref{fig_dnwd} or AM~Her in Figure~\ref{fig_amher}), suggesting that its spectrum may contain a larger proportional bulk WD contribution (compared to the accretion spots on the WD and/or the disk) than seen in the longer-period IPs. AE~Aqr is an unusual system in which the formation of the accretion disk is disrupted by a ``propeller'' mechanism caused by the rapidly rotating WD which ejects accreted matter from the system \citep{wynn1997}. Its FUV is similar to that of the other IPs.

Some of the lower inclination systems show blue continuum emission. In YY~Dra, \citet{hoard2005} used  the time-tag capabilities of \textit{FUSE} to create spectra phased over the spin period of the WD.  Their models of these spectra showed that the FUV emission predominantly originates from a $T_{WD} = 21,500$~K WD and two hot (T$_{spot}$ = 220,000~K) accretion spots on the WD surface separated by 180$^{\circ}$ in longitude.  Although they found the accretion disk to be negligible in YY~Dra, it may also contribute to the continuum in other systems.  Note that the continuum shape in YY~Dra is reminiscent of that of EX~Hya, which again points to a significant WD component in the latter.

V405~Aur is listed as a low-inclination target but its spectrum is more like that of the high-inclination systems, with a weak continuum and strong line emission. \citet{sing2004} measured the radial velocity curve of \ion{O}{6} 1032~\AA\ in V405~Aur and found a very low orbital radial velocity amplitude, $K=2.5\pm0.5$~km~s$^{-1}$, from which they inferred an orbital inclination of $<5^{\circ}$.   However, there are reasons to suspect that the \ion{O}{6} radial velocity curve is inaccurate. Its amplitude is significantly smaller than the amplitudes of 42--50~km~s$^{-1}$ seen in H$\alpha$, H$\beta$, and \ion{He}{2} $\lambda$4686 in optical spectra \citep{haberl1994}. The 2.5~km~s$^{-1}$ is also 1/12th a \textit{FUSE} resolution element, making its measurement challenging. Finally, because the authors tracked wavelength shifts using the Ly$\alpha$ geocoronal line, they were not sensitive to target shifts within the 40$\arcsec$ square aperture in the dispersion direction. If the \ion{O}{6} radial velocity curve is not tracking the WD's motion, the true WD radial velocity is likely to be higher than 2.5~km~s$^{-1}$. In that case, V405~Aur is likely to have a larger binary inclination and may fit within the normal pattern of morphological variation of the \textit{FUSE} IP spectra.

LS~Peg is an outlier:  the only IP with an absorption line spectrum. It may be mis-classified as an IP, as it does not show a X-ray periodicity corresponding to the WD spin period \citep{ramsay2008}. Ramsay et al.\ still classify LS~Peg as an IP however based on its strong, multi-component X-ray absorption spectrum. They attribute the lack of spin period modulation to a close alignment between the WD magnetic and spin axes.  

A few of the IPs show signs of anomalous line ratios, particularly in nitrogen vs.\ carbon emission line fluxes.  AE~Aqr is the extreme example, with very weak C and enhanced N emission lines.  TX~Col is similar.  At least 10 CVs, including AE~Aqr and TX~Col, have shown anomalously large \ion{N}{5}/\ion{C}{4} ratios in their UV spectra \citep{gansicke2003}.  As with the DN WD spectra that show such anomalies, the explanation is usually that CNO-processed material is present on the donor star and accreted through the disk onto the WD, as it is difficult to conceive of special photoionization conditions that could explain the line ratios across a wide variety of systems.

\subsection{Polars}

\begin{figure}
\plotone{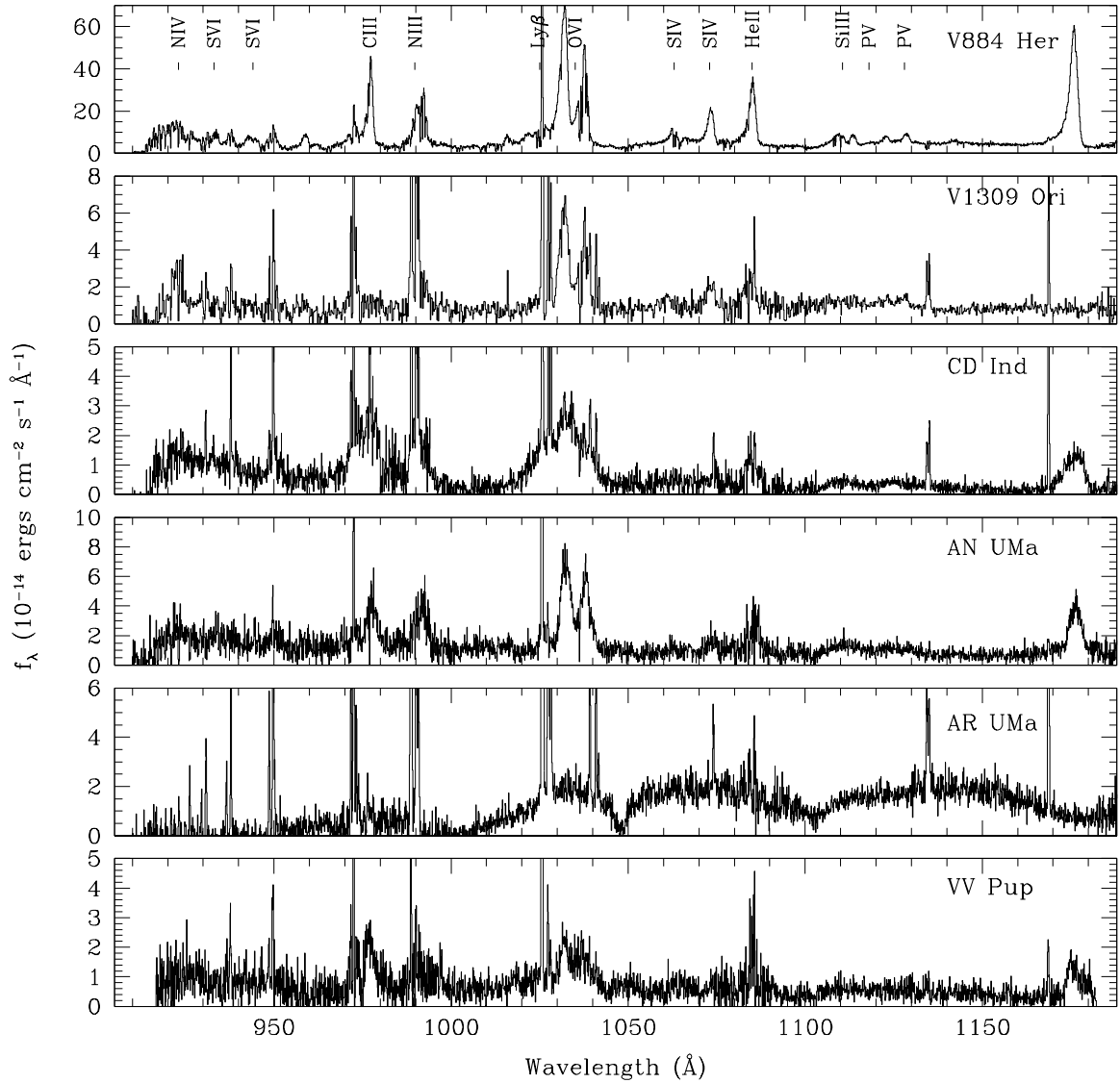}
\figcaption{The time-averaged spectra of six polars observed by \textit{FUSE}:  V884 Her, V1309 Ori, CD Ind, AN UMa, AR UMa, and VV Pup. A few of the prominent emission lines are labeled in the top panel.\label{fig_polar}}
\end{figure}

\textit{FUSE} observed 17 polars, designated ``NL AM" in Table~\ref{tab_list}.  Six of the targets (HS Cam, V895 Cen, EP Dra, BL Hyi, V347 Pav, EV UMa) were non-detections or marginal detections only.  In Fig~\ref{fig_polar}, we show some of the brighter polar spectra acquired by \textit{FUSE}. (Note, however, the flux scale on the figures: while these spectra are among the brightest of the observed polars, they are generally fainter than the other ``bright" CVs observed by \textit{FUSE}.)  Most of the polar spectra are characterized by strong emission lines in ionized species.  The lines and continuum are variable in time as the viewing angle of the accretion stream and impact point changes \citep{hutchings2002,hoard2002}. The anomalous N/C abundance ratio observed in some of the IPs is also seen in two of the polar spectra, BY~Cam and V1309~Ori, which in Figure~\ref{fig_polar} shows the weak \ion{C}{3} emission at 977~\AA\ and 1176~\AA\ as well as enhanced \ion{N}{4} $\lambda$923~\AA\ emission characteristic of this effect in the \textit{FUSE} waveband.  The N/C abundances in BY~Cam are analyzed in more detail by \citet{mouchet2003}.

In only a few cases --- AR~UMa, V834~Cen, and AM~Her --- can the underlying white dwarf spectrum be discerned.   The polar WDs are fainter and harder to detect than those in non-magnetic CVs: at any given orbital period, polars have cooler WDs than their non-magnetic counterparts, most likely due to their strong magnetic fields inhibiting magnetic braking, which will result in lower accretion rates, and hence less accretion heating \citep{araujo2005, townsley2009}.  In the highly magnetic polar, AR~UMa, the Zeeman splitting of the white dwarf's hydrogen lines can be modeled to obtain a 235 MG dipole magnetic field strength for the white dwarf \citep{hoard2004}.  The WD can best be seen in AM~Her when the accretion rate has dropped. Fig~\ref{fig_amher} shows the spectra of AM Her acquired on two different observing epochs.  In 2002, the spectrum shows the typical spectrum of an actively-accreting polar, while in 2000 the accretion activity had greatly diminished, revealing the 20,000~K white dwarf \citep{gaensicke2006}.  


\section{Conclusions} \label{sec_conclusion}

The \textit{FUSE} observations of CVs have provided a rich data set that allows observers to probe the physics of accretion and CV evolution among a diverse array of targets. Here, we summarize some of the key characteristics of the \textit{FUSE} CV spectra.

\begin{enumerate}

\item The DN quiescent spectra typically have spectra dominated by the white dwarf accretor, although many systems have strong emission line spectra wherein the WD component is obscured.  WD model fits to the spectra often show the need for additional components contributing to the FUV emission. These components have been interpreted as a hot, rapidly-spinning accretion belt on the surface of the WD, the boundary layer, the hot spot, and/or optically thin disk emission. Several of the systems show abundance anomalies on the WD indicative of previous CNO processing of the accreted material. The emission line dominated DN have FUV spectra that are challenging to model, as a result of our poor understanding of the properties of accretion disks at low accretion rates.

\item The spectra of DN in outburst show continuum emission from the steady-state accretion disk along with strong, broad spectral features from a disk outflow, varying from absorption to emission as the binary inclination increases. Some systems do not show strong wind lines but instead have narrow, low-velocity absorption lines which may originate in vertically extended material in the outer disk. Fits of steady-state accretion disk models show higher accretion rates for systems above the period gap than below, while excess emission above the predicted model fluxes near the Lyman limit suggest the presence of an additional continuum source in the FUV.

\item NL CVs are a broad class of objects and the \textit{FUSE} spectra are similarly broad and diverse in their properties. Several of the low-inclination systems show spectra with broad, blueshifted absorption lines characteristic of strong disk winds.  However, many of the other NLs have relatively narrow absorption line spectra centered on or near their rest velocities which may indicate the presence of vertically-extended disk components but no strong outflow. The higher inclination NLs have emission line spectra from the disk wind as well as possible disk chromospheric components seen in absorption. The continuum emission in NLs is dominated by the accretion disk, although steady-state disk models often deviate from the observations, particularly at the bluest wavelengths.

\item The \textit{FUSE} spectra of IPs are characterized by strong emission lines and blue continuum emission. Model fits shows that the continuum emission comes from the bulk of the WD, the hot spots at the base of the magnetically-channeled accretion columns, and possibly the disk. Some of the IPs shows enhanced N and weak C emission lines indicative of CNO processed material in the accretion stream.

\item The \textit{FUSE} spectra of polars are relatively faint. They show strong emission lines and either blue continua or a continuum consistent with a cool WD, depending on the accretion rate. Two of the polars show the enhanced N and weak C seen in a few of the IP spectra. 

\end{enumerate}

\acknowledgements Our thanks to Louise Hartley for providing the time-averaged spectrum of Z~Cam. The authors also wish to thank NASA and the \textit{FUSE} GO program for their financial support for this program. All of the data presented in this paper were obtained from the Multimission Archive at the Space Telescope Science Institute (MAST). STScI is operated by the Association of Universities for Research in Astronomy, Inc., under NASA contract NAS5-26555. Support for MAST for non-HST data is provided by the NASA Office of Space Science via grant NNX09AF08G and by other grants and contracts. We acknowledge with thanks the variable star observations from the AAVSO International Database contributed by observers worldwide and used in this research.

Facilities: \facility{\textit{FUSE}},\facility{AAVSO}



\clearpage
\begin{deluxetable}{llcccllccll}
\rotate 
\tabletypesize{\scriptsize} 
\tablecaption{Cataclysmic Variables Observed by \textit{FUSE}\label{tab_list}} 
\tablewidth{0pt} \tablecolumns{11}
\tablehead{ \colhead{Object} & \colhead{CV Type\tablenotemark{a}} & \colhead{P$_{orb}$} & \colhead{$i$} &
\colhead{Data ID} & \colhead{Start Date} & \colhead{T$_{exp}$} &
\colhead{TTAG/HIST} & \colhead{Data Quality\tablenotemark{b}} & \colhead{Comments}  & \colhead{References} 
\\ & & \colhead{(hr)} & \colhead{($^{\circ}$)} & & \colhead{(UT)}  & \colhead{(s)} \\ } 
\startdata
IW And & DN ZC & & & D9130801 & 2003 Dec 9 & 17635 & TTAG & 3 \\ 
RX And & DN ZC & 5.04 & 51 & B0700101 & 2001 Sept 2 & 8211 & TTAG & 4 & Outburst \\                      
             & & & & C0530301 & 2002 Aug 25 & 16729 & TTAG & 4 & Outburst rise \\
DT Aps & DN UG & & & G9251101 & 2006 Apr 28 & 8037 & TTAG & 3 \\
	& & & & G9251102 & 2006 May 3 & 68216 & TTAG & 3 & & Go2009b \\      
AE Aqr & NL IP & 9.88 & 58 & B0340101 & 2001 Jul 17 & 42935  & TTAG & 4 \\ 	               
UU Aql & DN UG & 3.37 & & C1100301 & 2004 May 16 & 17051 & TTAG & 3 &  & Si2007 \\
V794 Aql & NL VY & 5.52 & 39 & D1440101 & 2004 May 13 & 13048 & TTAG & 4 &  & Go2007  \\
V1432 Aql & NL AM & 3.37 & & D1090101 & 2004 May 14 & 12857 & TTAG & 3 \\
          &       &      & & D1090102 & 2004 May 17 & 10952 & TTAG & 3 \\
V433 Ara & DN SU & & & G9250901 & 2007 Feb 18 & 7009 & TTAG & 3 &  & Go2009b \\ 
V499 Ara & DN UG & & & G9251001 & 2006 Jun 22 & 17988 & TTAG & 1  & & Go2009b \\
V663 Ara & DN SU & & & G9250801 & 2006 Apr 24 & 9774 & TTAG & 1 & & Go2009b \\ 
	& & & & G9250802 & 2007 Apr 20 & 4822 & TTAG & 1 \\
TT Ari & NL VY & 3.30 & 30\tablenotemark{c} & P1840501 & 2004 Jan 17 & 5925 & TTAG & 4 & & Hu2007  \\
SS Aur & DN UG & 4.39 & 38 & C1100201 & 2002 Feb 13 & 14486 & TTAG & 4 &  &
Si2004a \\
V405 Aur & NL IP & 4.15 & $<$5 & Z9100701 & 2002 Oct 15 & 7916 & TTAG & 2 \\
	   & & & & D0800101 & 2003 Oct 1 & 22686 & TTAG & 3 &  & Sg2004 \\
	   BY Cam & NL AM & 3.35 & $>$40 & A0240107 & 2000 Jan 10 & 12631 & TTAG & 4 &  & Mo2003 \\
       &       &      & & A0240108 & 2000 Jan 16 & 22124 & TTAG & 4 \\
       & & & & U1020303 & 2006 Aug 19 & 26517 & TTAG & 3 \\
       & & & & U1020304 & 2007 Feb 02 & 13557 & TTAG & 2 \\
BZ Cam & NL VY & 3.69 & $<$40\tablenotemark{c} & D9050101 & 2003 Dec 1 & 13187 & TTAG & 2 \\
	& & & & U1020901 & 2005 Nov 01 & 0 & TTAG & 1 \\ 
HS Cam & NL AM & 1.64 & 79 & Z9102001 & 2003 Jan 30 & 73827 & TTAG & 2  \\
	& & & & U1021702 & 2007 Feb 6 & 52288 & TTAG & 2 \\	
Z Cam & DN ZC & 6.96 & 57 & C0680101 & 2002 Feb 9 & 24436 & HIST & 3 &  & Ha2004 \\ 
AM CVn & NL AC & 0.29 & & C0080101 & 2004 Dec 12 & 13011 & TTAG & 3 \\
QU Car & NL & 10.9 & $<$60 & Z9103601 & 2002 Jun 28 & 9265 & TTAG &  3 \\
       &    &      &       & D1560101 & 2003 Apr 22 & 28756 & TTAG & 3 \\
       &    &      &       & D1560102 & 2003 Apr 23 & 10763 & TTAG & 4 \\
V436 Car & NL IP & 4.20 & & Z9100201 & 2003 Dec 25 & 10944 & TTAG & 4 \\
         & & & & D9130301 & 2004 Dec 10 & 7054 & TTAG & 3 \\
         & & & & U1071501 & 2007 Jul 05 & 7008 & TTAG & 3 \\       
AM Cas & DN ZC & 3.96 & & G9251402 & 2006 Oct 19 & 15920 & TTAG & 3 & & Go2009b \\
V592 Cas & NL UX & 2.76 & 28\tablenotemark{c} & D1140101 & 2003 Aug 5 & 23751 & TTAG & 4 &  & Pr2004 \\
& & &  & D1140102 & 2003 Aug 7 & 19872 & TTAG & 3 \\
& & &  & D1140103 & 2003 Aug 8 & 14283 & TTAG & 4 \\
V709 Cas & NL IP & 5.4 & & Z9100401 & 2002 Oct 25 & 10524  & TTAG & 2 \\
	& & & & Z9100501 & 2002 Dec 15 & 28050 & TTAG  & 2 \\
	& & & & U1011201 & 2006 Oct 29 & 27854 & TTAG & 2 \\
	& & & & U1011202 & 2006 Oct 29 & 12814 & TTAG & 2 \\
	& & & & U1011203 & 2006 Oct 30 & 22313 & TTAG & 2 \\
BV Cen & DN UG & 14.6 & 53 & D1450301 & 2003 Apr 13 & 27896 & TTAG & 3 &  & Si2007 \\
V803 Cen & NL AC & 0.45 & 13.5 & C0600101 & 2003 Apr 10 & 26042 & TTAG & 4\\
V834 Cen & NL AM & 1.69 & 50 & Z9105010 & 2002 Jun 23 & 25512 & TTAG & 4 \\
V895 Cen & NL AM & 4.77 & 80 & C1620201 & 2003 Apr 10 & 25097 & TTAG & 1 \\
WX Cen & NL? & 10.0 & & D9131001 & 2003 Jun 12 & 6573 & TTAG & 3 \\
WW Cet & DN & 4.22 & 54 & D1450401 & 2003 Jul 26 & 14803 & TTAG & 4 &  & Go2006 \\
TX Col & NL IP & 5.72 & 25 & D9050201 & 2003 Dec 31 & 3525 & TTAG & 3 \\
EM Cyg & DN ZC & 6.98 & 67\tablenotemark{c} & C0100101 & 2002 Sep 5 & 8885 & TTAG & 4 &  & Go2009 \\
EY Cyg & DN UG & 5.24 & 14 & D1450101 & 2003 Jul 16 & 18509 & TTAG & 2 &  & Si2004b \\
SS Cyg & DN UG & 6.6 & 37 & A1260207 & 2000 Nov 2  & 5607 & HIST & 3 & Outburst peak \\
                    & & & & A1260208 & 2000 Nov 7  & 7547 & HIST & 3 & Outburst decline \\
                    & & & & A1260206 & 2000 Nov 10 & 7482 & HIST & 3 & Outburst decline \\
                    & & & & A1260210 & 2000 Nov 14 & 11458 & TTAG & 4 & Outburst end \\
                    & & & & P2420101 & 2001 Sept 4 & 1883 & TTAG & 4 & Quiescence & Si2010 \\
                    & & & & C0530201 & 2002 Jul 2 & 5153\tablenotemark{d} & TTAG & 3 & Outburst \\
                    & & & & C0530202 & 2002 Oct 26 & 9522 & TTAG & 3 & Quiescence \\
V751 Cyg & NL VY & 6.0 & & D9130701 & 2003 Jun 22 & 8563 & TTAG & 2 \\
BB Dor & NL & 3.6 & & H9030301 & 2007 Jul 11 & 1567 & TTAG & 4 &  & Go2008 \\
AB Dra & DN ZC & 3.65 & & D9050301 & 2003 Dec 16 & 15574 & TTAG & 4 \\
				& & & & E9890401 & 2007 Mar 8 & 6320 & TTAG & 4 \\
				& & & & E9890402 & 2007 Mar 9 & 1038 & TTAG & 3 \\
				& & & & E9890403 & 2007 Mar 9 & 12917 & TTAG & 4 \\
				& & & & E9890404 & 2007 Mar 11 & 600 & TTAG & 3\\
CQ Dra & ? & 3.97 & & D9050401 & 2004 Jan 9 & 36719 & TTAG & 4 & Symbiotic? \\
		   & & & & U1031201 & 2005 Dec 11 & 6745 & TTAG & 1 \\
EP Dra & NL AM & 1.74 & 80 & Z9102401 & 2002 Jul 14 & 14744 & TTAG & 1 \\		   
ES Dra & DN? & 4.24 & & G9251601 & 2006 Nov 19 & 24534 & TTAG & 3 & & Go2009b \\
EX Dra & DN UG & 5.04 & 84.2 & D9050501 & 2004 Feb 16 & 22099 & TTAG & 3 \\
& & & & D9050502 & 2004 Feb 17 & 19521 & TTAG & 3 \\
YY Dra & NL IP & 3.97\tablenotemark{c} & 45\tablenotemark{c} & C0410201 &
2002 Jan 30 & 53556 & TTAG & 4 &  &  Ho2005 \\ 
U Gem & DN UG & 4.25 & 69.7 & A1260112 & 2000 Mar 5 & 2883 & HIST & 4 & Outburst peak & Fr2001, Lo2006 \\
                            & & & & A1260102 & 2000 Mar 7 & 6248 & HIST & 3 & Outburst plateau \\
                            & & & & A1260103 & 2000 Mar 9 & 7868 & HIST & 3 & Outburst plateau \\
                            & & & & A1260105 & 2000 Mar 17 & 13052 & TTAG & 3 & Outburst end \\
                            & & & & P1540203 & 2001 Feb 22 & 12990 & TTAG & 4 & Quiescence &  \\
RZ Gru & NL UX & 10.0 & & D9050601 & 2003 Sep 13 & 6118 & TTAG & 4 \\
AM Her & NL AM & 3.09 & 60 & P1840601 & 2000 May 12 & 5619 & TTAG & 4 &  & Hu2002, Ga2006 \\
       &       &      &    & Z0060101 & 2002 May 11 & 46851 & TTAG & 4 \\
       &       &      &    & C0530503 & 2002 Sep 8 & 5407 & TTAG & 4 \\
       &       &      &    & C0530504 & 2002 Sep 8 & 8471 & TTAG & 4 \\
       &       &      &    & C0530505 & 2002 Sep 8 & 7243 & TTAG & 4 \\
DQ Her & NL IP & 4.65 & 86.5 & D9130501 & 2003 Jun 30 & 7816 & TTAG & 4 \\       
V795 Her & NL & 2.60 & & P1840701 & 2000 Aug 11 & 4830 & TTAG & 4 \\
V884 Her & NL AM & 1.88 & & B1010101 & 2001 Jul 25 & 34692 & TTAG &  4 \\
EX Hya & NL IP & 1.64 & 78 & A0840101 & 2000 May 18 & 5674 & TTAG & 4 \\
                     & & & & A0840102 & 2000 May 19 & 5458 & TTAG & 4 \\
                     & & & & A0840103 & 2000 May 19 & 4456 & TTAG & 4  \\
                     & & & & A0840104 & 2000 May 20 & 6100 & TTAG & 4 \\
                     & & & & A0840105 & 2000 May 21 & 5810 & TTAG & 4 \\
VW Hyi & DN SU & 1.78 & 60 & B0700201 & 2001 Aug 18 & 17435 & TTAG & 4 & Quiescence & Go2004b, Go2005, Lo2009 \\
	& & & & C0530401 & 2002 Jul 28 & 18554 & TTAG & 3 & Quiescence \\
	& & & & E1140101 & 2004 Jul 30 & 5984 & TTAG & 4 & Superoutburst decline \\
	& & & & E1140102 & 2004 Jul 31 & 1080 & HIST & 4 & Superoutburst end \\
	& & & & E1140103 & 2004 Aug 1 & 3482 & TTAG & 4 & Quiescence \\
	& & & & E1140104 & 2004 Aug 2 & 5782 & TTAG & 4 & Quiescence \\
	& & & & E1140105 & 2004 Aug 3 & 4869 & TTAG & 4 & Quiescence \\	
	& & & & E1140106 & 2004 Aug 4 & 8454 & TTAG & 4 & Quiescence \\
	& & & & E1140107 & 2004 Aug 5 & 26009 & TTAG & 3 & Quiescence \\
	& & & & E1140108 & 2004 Aug 7 & 23225 & TTAG & 4 & Quiescence \\					
	& & & & E1140109 & 2004 Aug 10 & 24465 & TTAG & 4 & Quiescence \\
	& & & & E1140110 & 2004 Aug 13 & 26790 & TTAG & 4 & Quiescence \\
	& & & & E1140111 & 2004 Aug 16 & 19853 & TTAG & 4 & Outburst \\
	& & & & E1140112 & 2004 Aug 19 & 27592 & TTAG & 4 & Quiescence \\
	& & & & E1140113 & 2004 Aug 30 & 22336 & TTAG & 4 & Quiescence \\
	& & & & U1064401 & 2006 Jul 23 & 4463 & TTAG & 4 & Quiescence \\
BL Hyi & NL AM & 1.89 & 70 & Z9100801 & 2002 Sept 29 & 10373 & TTAG & 2 \\
	& & & & U1061901 & 2006 May 30 & 12179 & TTAG & 1 \\	
	& & & & U1061902 & 2006 July 30 & 12272 & TTAG & 1 \\
	& & & & U1061903 & 2006 Sept 30 & 10385 & TTAG & 1 \\
	& & & & U1061904 & 2006 Nov 27 & 7257 & TTAG & 1 \\
WX Hyi & DN SU & 1.80 & 40 & D9050701 & 2003 Nov 6 & 22213 & TTAG & 3 & Quiescence \\
	& & & & S7012101 & 2005 Apr 16 & 0 & TTAG & 1 \\
	& & & & S7012102 & 2005 Apr 16 & 18302 & TTAG & 3 \\
		& & & & S7012103 & 2005 Apr 17 & 9314 & TTAG & 3 \\
	& & & & U1062001 & 2006 May 31 & 24008 & TTAG & 3 \\
	& & & & U1062002 & 2006 June 1 & 40624 & TTAG & 3 \\
CD Ind & NL AM & 1.85 & & Z9101701 & 2003 May 27 & 31191 & TTAG & 3 \\
BH Lyn & NL VY & 3.74 & & D9131501 & 2003 Apr 2 & 5506 & TTAG  & 2 \\
BP Lyn & NL UX & 3.67 & & D9050801 & 2003 Mar 28 & 5476 & TTAG  & 4 \\
MV Lyr & NL VY & 3.19 & 12 & C0410301 & 2002 Jul 7 & 11212 & TTAG & 4 & Low state & Ho2004 \\
                    & & & & D9050901 & 2003 May 6 & 7637 & TTAG & 3 & High state & Go2011 \\
AH Men & NL & 2.95 & & D9051001 & 2003 Apr 30 & 23964 & TTAG & 3 \\
       &    &      & & D9051002 & 2004 Sep 22 & 10694 & TTAG & 3 \\
AQ Men & DN UG &3.40 & & G9250201 & 2006 Nov 22 & 22661 & TTAG & 2 & & Go2009b  \\       
HP Nor & DN ZC & & & G9250601 & 2007 Apr 13 & 3730 & TTAG & 2 & & Go2009b \\
IK Nor & DN UG & & & G9250701 & 2007 Apr 13 & 7757 & TTAG & 1 & & Go2009b \\
V1309 Ori & NL AM & 7.98 & 78 & A0240201 & 2000 Nov 4 & 11294 & TTAG & 4 \\
BD Pav & DN UG & 4.30 & 71 & D9130101 & 2003 May 31 & 10714 & TTAG & 1 \\
V345 Pav & NL UX & 4.75 & & D9131101 & 2003 May 26 & 24709 & TTAG & 4 \\
	& & & & U1095002 & 2006 Aug 16 & 13026 & TTAG & 3 \\
V347 Pav & NL AM & 1.50 & 80 & Z9101601 & 2002 Aug 6 & 22050 & TTAG & 2 \\
         &       &      &    & Z9101602 & 2002 Aug 7 & 23044 & TTAG & 2 \\
         &       &      &    & Z9101603 & 2002 Aug 13 & 6271 & TTAG & 2 \\
         & & & & U1094302 & 2006 Jun 24 & 18694 & TTAG & 1 \\
         & & & & U1094304 & 2007 Jun 17 & 1563 & TTAG & 2 \\
LS Peg & NL IP & 3.6 &30  & B0480101 & 2001 Jul 10 & 13270 & TTAG & 4 \\
RU Peg & DN UG & 8.99 & 33 & C1100101 & 2002 Jul 4 & 3026 & TTAG & 4 &  & Si2004a \\
FO Per & DN UG & & & G9251501 & 2007 Feb 11 & 2716 & TTAG & 3 & & Go2009b \\
FY Per & NL UX & 6.20 & & D9130401 & 2004 Jan 26 & 7395 & TTAG & 3  \\
KT Per & DN ZC & 3.90 & & D9051101 & 2003 Oct 11 & 28474 & TTAG & 3 \\
CM Phe & NL & 6.45 & & D9131801 & 2003 Sep 16 & 5684 & TTAG & 1 \\
TW Pic & NL IP & 6.06 & & D9051201 & 2003 Sep 7 & 20492 & TTAG & 3 \\
VV Pup & NL AM & 1.67 & 75 & B0470201 & 2001 Apr 5 & 17667 & TTAG & 3 &  & Ho2002 \\
V347 Pup & NL IP & 5.57 & 84 & D9051301 & 2003 Mar 9 & 18192 & TTAG & 4 \\
V Sge & NL & 12.3 & 90 & B0430101 & 2001 Jul 11 & 13705 & TTAG & 3 \\
      &    &      &    & B0430102 & 2001 Nov 8 & 3599 & TTAG & 4 \\
      &    &      &    & B0430103 & 2001 Nov 9 & 3589 & TTAG & 4 \\
      & & & & B0430104 & 2004 May 30 & 7459 & TTAG & 4 \\
WZ Sge & DN SU & 1.36 & 75 & Z0030101 & 2001 Jul 30 & 3180 & HIST & 3 &Outburst peak & Lo2003, Go2004a \\
       & & & & Z0030102 & 2001 Sep 7 & 8276 & HIST & 3 & Rebrightening & Lo2003 \\
       & & & & Z0030103 & 2001 Sep 29 & 9051 & TTAG & 4 & Outburst decline \\
       & & & & Z0030104 & 2001 Nov 3 & 3155 & TTAG & 3 & Outburst decline \\
       & & & & Z0030105 & 2001 Nov 5 & 2634 & TTAG & 4 & Outburst decline \\
       & & & & Z0030106 & 2001 Nov 7 & 5940 & TTAG & 4 & Outburst decline \\
       & & & & Z0030107 & 2001 Nov 8 & 3682 & TTAG & 4 & Outburst decline \\
V3885 Sgr & NL UX & 5.19 & $<$50 & P1870101 & 2000 May 24 & 12392 & TTAG & 4 & & Li2009 \\ 
          &       &      &       & Z9104601 & 2002 Aug 10 & 1714 & HIST & 4\\
          &       &      &       & D9051501 & 2003 Sep 21 & 8320 & TTAG & 4 \\
          &       &      &       & D9051502 & 2003 Sep 22 & 7045 & TTAG & 4 \\
RW Sex & NL UX & 5.88 & 34 & B1040101 & 2001 May 13 & 25599 & TTAG & 4 &  & Pr2003, Li2010 \\
QS Tel & NL AM & 2.33 & & C1610201 & 2002 Jan 6 & 22175 & TTAG & 3 \\
EF Tuc & DN UG & 3.48  & & E9890701 & 2004 Aug 22 & 14294 & TTAG & 4 \\
	& & & & U1060101 & 2006 Oct 3 & 5727 & TTAG & 3 \\
VW Tuc & DN UG & & & G9250101 & 2006 Jun 11 & 7529 & TTAG & 1 & & Go2009b \\ 
EK TrA & DN SU & 1.53 & 58 & Z9104301 & 2002 Jun 24 & 33491 & TTAG & 4 & & Go2008 \\
       	&  & & & Z9104302 & 2002 Jun 24 & 32979 & TTAG & 4 \\
       	& & & & U1091601 & 2006 Mar 01 & 5386 & TTAG & 3 \\
	& & & & U1091602 & 2007 Feb 17 & 7583 & TTAG & 3 \\
	& & & & U1091603 & 2007 Feb 17 & 6394 & TTAG & 3 \\
	  & & & & U1091604 & 2007 Feb 18 & 3159 & TTAG & 2 \\
	  & & & & U1091605 & 2007 Apr 16 & 432 & TTAG & 1 \\
	  & & & & U1091606 & 2007 Apr 18 & 4501 & TTAG & 2 \\
AN UMa & NL AM & 1.91 & 40--60\tablenotemark{c} & Z9102101 & 2003 May 20 & 9640 & TTAG & 4 \\
AR UMa & NL AM & 1.93 & 50 & B0470101 & 2003 Jan 21 & 10644 & TTAG & 3 & &  Ho2004 \\
       &       &      & & B0470102 & 2003 Jan 21 & 11161 & TTAG & 3  \\
EV UMa & NL AM & 1.33 & & Z9102201 & 2004 Mar 13 & 6331 & TTAG & 2 \\
CH UMa & DN UG & 8.23 & 21 & D1450201 & 2003 Apr 2 & 17302 & TTAG & 4 &  & Si2007 \\
DW UMa & NL UX & 3.28 & 80 & B0480201 & 2001 Nov 7 & 18493 & TTAG & 4 &  & Ho2003 \\ 
ER UMa & DN SU & 1.53 & & D9051401 & 2004 Jan 15 & 30625 & TTAG & 3 \\
SU UMa & DN SU & 1.83 &    & E9891201 & 2004 Nov 9 & 11157 & TTAG & 3 \\
SW UMa & DN SU & 1.36 & 45 & B0740101 & 2001 Nov 6 & 26202 & TTAG & 2 &  & Po2004 \\
UX UMa & NL UX & 4.72 & 71 & B0820101 & 2001 Mar 23 & 22828 & TTAG & 4 &  & Fr2003, Li2008 \\
       &       &      &    & B0820102 & 2001 Mar 24 & 21179 & TTAG & 4 \\
       &       &      &    & Z9104701 & 2003 Jan 16 & 11433 & TTAG & 4 \\
IX Vel & NL UX & 4.65 & 60 & Q1120101 & 2000 Apr 15 & 6089 & HIST & 4 &  & Li2007b \\
       &       &      &    & P2042501 & 2002 Mar 12 & 40526 & HIST & 3 & MDRS \\
       &       &      &    & P2042502 & 2002 May 5 & 8865 & HIST & 3 & MDRS \\
       &       &      &    & P2042503 & 2003 Dec 16 & 7392 & HIST & 3 & MDRS \\
       &       &      &    & P2042504 & 2003 Dec 17 & 11107 & HIST & 3 & MDRS \\
       &       &      &    & P2042505 & 2003 Dec 18 & 11590 & HIST & 3 & MDRS \\
HS0551+7241 & NL & 0.143 & & Z9100601 & 2002 Oct 14 & 10266 & TTAG & 1 & (LS Cam) \\
	& & & & U1020402 & 2006 Aug 18 & 15812 & TTAG & 1 \\
	& & & & U1020403 & 2006 Oct 13 & 8001 & TTAG & 1 \\
	& & & & U1020404 & 2006 Oct 15 & 11547 & TTAG & 1 \\
HS1857+7127 & DN & & & E9891301 & 2004 Oct 11 & 9292 & TTAG & 3 \\
	& & & & U1041102 & 2006 Sep 13 & 9146 & TTAG & 2 \\
	& & & & U1041103 & 2007 Mar 9 & 16966 & TTAG & 3 & Outburst? \\
NSV-10934 & DN SU & & & G9251201 & 2006 Jun 28 & 11956 & TTAG & 3 & & Go2009b \\
 & & & & G9251202 & 2006 Jun 30 & 14177  & TTAG & 2 \\
P831-57 & NL & & & H9030201 & 2007 Jul 09 & 10029 & TTAG & 3 \\ 
SDSS015543+0028 & NL & & & D0420101 & 2004 Aug 31 & 19258 & TTAG & 1 & NL AM? \\
RXJ0625+7334 & NL\tablenotemark{e} & 4.72 & & E9890801 & 2004 Nov 6 & 12555 & TTAG & 3 & (MU Cam); NL IP? \\
	& & & & U1020803 & 2006 Oct 15 & 7860 & TTAG & 3 \\
	& & & & U1020804 & 2006 Dec 8 & 7019 & TTAG & 3 \\
RXJ1548--4528 & NL IP & 9.86 & & Z9100301 & 2002 Jun 22 & 22725 & TTAG & 3 \\
SDSS080908+3841 & NL & & & D0420201 & 2003 Mar 28 & 8447 & TTAG & 3  & & Li2007 \\
                & & & & D0420202 & 2004 Mar 16 & 10392 & TTAG & 3 \\
0932+4950 & NL & 10.04 & & D9130201 & 2004 Mar 17 & 9836 & TTAG & 2 \\
\enddata
\tablecomments{CV types, orbital periods, and inclinations taken from
Ritter \& Kolb 2003 and Downes et al.\ 2001 except where noted.  References:
Froning et al\ 2003 (Fr2003); Froning et al. 2001 (Fr2001); G{\"a}nsicke et al. 2006 (Ga2006); Godon \& Sion 2011 (Go2011); Godon et al.\ 2009b (Go2009b); Godon et al.\ 2009 (Go2009); Godon et  al.\ 2008 (Go2008); Godon et al.\ 2007 (Go2007); Godon et al.\ 2006 (Go2006); Godon et al.\ 2005 (Go2005); Godon et al.\ 2004a (Go2004a); Godon et al.\ 2004b (Go2004b); Hartley et al.\ 2005 (Ha2005); Hoard et al. 2004  (Ho2004);
Hoard et al.\ 2003, AJ (Ho2003);  Hoard et al.\ 2002 (Ho2002);  Hutchings \& Cowley 2007 (Hu2007); Hutchings et al.\ 2002 (Hu2002); Linnell et al.\ 2010 (Li2010); Linnell et al. 2009 (Li2009); Linnell et al. 2008 (Li2008); Linnell et al.\ 2007 (Li2007); Linnell et al.\ 2007b (Li2007b); Long et al.\ 2009 (Lo2009); Long et al.\ 2006 (Lo2006); Long et al.\ 2003 (Lo2003); Mouchet et al.\ 2003 (Mo2003); Povich et al.\ 2004
(Po2004); Prinja et al.\ 2004 (Pr2004); Prinja et al.\ 2003 (Pr2003); Sing et al.\ 2004 (Sg2004); Sion et al.\ 2010 (Si2010); Sion et al.\ 2007 (Si2007); Sion et al.\ 2004a (Si2004a); Sion et al.\ 2004b (Si2004b).}
\tablenotetext{a}{The CV type abbreviations are as follows:  ``DN" = dwarf nova; ``NL" = novalike;  ``AC" = AM CVn subtype of NL; ``AM" = AM Her (or ``polar") subtype of NL;``IP" = intermediate polar subtype of NL; ``SU" = SU  UMa subtype of DN; ``UG" = U Gem subtype of DN; ``UX" = UX UMa subtype of NL; ``VY" = VY Scl subtype of NL;  ``ZC" = Z Cam subtype of DN.}
\tablenotetext{b}{Data quality flag indicates the quality of the observation and the pipeline processing. A rank of 4 indicates that there were no serious problems with the observation and the source was well detected; 3 indicates a good source detection but one or more problems with the observation or data processing; 2 indicates that the source was detected but the data quality is marginal; and 1 indicates that the source was not detected.}
\tablenotetext{c}{Inclinations from: Bonnet-Bidaud et al. 1996 (AN UMa); Shafter et al.\ 1985 (TT Ari); Prinja et al.\ 2000 (BZ Cam); Huber, Howell, \& Ciardi 1998 (V592 Cas); North et al. 2000 (EM Cyg); orbital period and inclination from Haswell et al.\ 1997 (YY Dra)}
\tablenotetext{d}{Recorded exposure time down from actual on-target time of 8387 sec due to data buffer overflows in time-tag observing mode.}
\tablenotetext{e}{Araujo-Betancor et al.\ 2003 classify this as a NL IP magnetic CV.}
\end{deluxetable}

\begin{deluxetable}{cccccccc}
\tablecaption{Intrinsic Spectral Features in the FUV Spectra of CVs\label{tab_lines}}
\tablewidth{0pt}
\tablecolumns{8}
\tablehead{\colhead{Ion} & \colhead{$\lambda$ (\AA)} & & \colhead{Ion} & \colhead{$\lambda$ (\AA)} & & \colhead{Ion} & \colhead{$\lambda$ (\AA)}}
\startdata
N~\sc{iv}$\: \star$ & 922--924 & & N~\sc{iii}$\: \star$ & 1006.0 & & Si~\sc{iii}$\: \star$ & 1113.1 \\
H~\sc{i} & 923.2 & & C~\sc{ii}$\: \star$ & 1010.1 & & P~\sc{v} & 1118.0 \\
H~\sc{i} & 926.2  & & S~\sc{iii} & 1012.5 & & Fe~\sc{iii} & 1122.5 \\
H~\sc{i} & 930.7   & & S~\sc{iii} & 1015.5 & & Si~\sc{iv}$\: \star$ & 1122.5 \\
S~\sc{vi} & 933.4 &&  S~\sc{iii} & 1021.1 & & S~\sc{ii}$\: \star$ & 1124.4 \\
H~\sc{i} & 937.8   & & L$\beta$  & 1025.7 & & Fe~\sc{iii} & 1124.9 \\
S~\sc{vi} & 944.5 & & O~\sc{vi} & 1031.9 & & C~\sc{iii}$\: \star$ & 1125.6 \\
H~\sc{i} & 949.7  & & O~\sc{vi} & 1037.6 & & P~\sc{v} & 1128.0 \\
N~\sc{iv}$\: \star$ & 955.3 & & S~\sc{iv} & 1062.7 & & Si~\sc{iv}$\: \star$ & 1128.3 \\
He~\sc{ii}$\: \star$ & 958.7 & & C~\sc{ii}$\: \star$ & 1066.0 & & C~\sc{ii}$\: \star$ & 1141.6 \\
H~\sc{i} & 972.5 & & Si~\sc{iv}$\: \star$ & 1066.6 & & Si~\sc{iii}$\: \star$ & 1144.3 \\
C~\sc{iii} & 977.0 & & S~\sc{iv} & 1073.0 & & O~\sc{iii}$\: \star$ & 1149.6 \\
N~\sc{iii}$\: \star$ & 979.8 & & S~\sc{iii}$\: \star$ & 1077.2 & & O~\sc{iii}$\: \star$ & 1150.9 \\
N~\sc{iii} &  989.8 & & He~\sc{ii}$\: \star$ & 1084.9 & & O~\sc{iii}$\: \star$ & 1153.8 \\
He~\sc{ii}$\: \star$ & 992.4 & &  S~\sc{iv}$\: \star$ & 1098.9 & & C~\sc{iv}$\: \star$ & 1169.0 \\
Si~\sc{iii}$\: \star$ & 993.5 & & Si~\sc{iii}$\: \star$ & 1108.4 & & C~\sc{iii}$\: \star$ & 1175.3 \\
Si~\sc{iii}$\: \star$ & 997.4 & & Si~\sc{iii}$\: \star$ & 1110.0 \\
\enddata
\tablecomments{Wavelengths are taken from \citet{morton1991} and the Atomic Line List v.2.04 (\url{http://www.pa.uky.edu/~peter/atomic/}).  Lines marked with a $\star$ are excited state transitions.}
\end{deluxetable}

\begin{deluxetable}{cccccccc}
\tablecaption{Prominent Intersteller Lines in the FUV Spectra of CVs\label{tab_ism}}
\tablewidth{0pt}
\tablecolumns{6}
\tablehead{
\colhead{Ion} & \colhead{$\lambda$ (\AA)}  & & \colhead{Ion} & \colhead{$\lambda$ (\AA)} & & \colhead{Ion} & \colhead{$\lambda$ (\AA)}}
\startdata
H \sc{i} & 913.826  & & H \sc{i} & 937.803  & & O \sc{i} & 988.578  \\
H \sc{i} & 914.039  & & O \sc{i} & 948.686  & &            & 988.655  \\
H \sc{i} & 914.286  & & H \sc{i} & 949.743  &  &         & 988.773  \\
H \sc{i} & 914.576  & & O \sc{i} & 950.885  & & N \sc{iii} & 989.799  \\
H \sc{i} & 914.919  & & N \sc{i} & 952.303  & & Si \sc{ii} & 989.873  \\
H \sc{i} & 915.329  & &             & 952.415  & & Si \sc{ii} & 1020.699  \\
H \sc{i} & 915.824  & & N \sc{i} & 953.415  & & H \sc{i} & 1025.722  \\
H \sc{i} & 915.824 & &           & 953.655  & & O \sc{i} & 1025.762  \\
H \sc{i} & 916.429  & &           & 953.970  & & C \sc{ii} & 1036.337  \\
H \sc{i} & 917.181  &  &          & 954.104  & & O \sc{i} & 1039.230  \\
H \sc{i} & 918.129  & & P \sc{ii} & 963.801  & &  Ar \sc{i} & 1048.220 \\
H \sc{i} & 919.351  & & P \sc{ii} & 963.801  & & N \sc{ii} & 1083.990  \\
H \sc{i} & 920.963  & & N \sc{i} & 963.990  & & N \sc{i} & 1134.165  \\
H \sc{i} & 923.150  &  &          & 964.626  &       &     & 1134.415  \\
O \sc{i} & 924.950  &  &          & 965.041  &         &  & 1134.980  \\
H \sc{i} & 926.226  & & O \sc{i} & 971.738  &         &   & 1134.980  \\
O \sc{i} & 929.517  & & H \sc{i} & 972.537  & & Fe \sc{i} & 1144.946  \\
H \sc{i} & 930.748  & & O \sc{i} & 976.448  \\
O \sc{i} & 936.629  & & C \sc{iii} & 977.020  \\
\enddata
\tablecomments{Wavelengths for the lines are taken from \citet{morton1991}.}
\end{deluxetable}



\figsetstart
\figsetnum{1}
\figsettitle{Non-Magnetic CVs Observed by \textit{FUSE}}

\figsetgrpstart
\figsetgrpnum{1.1}
\figsetgrptitle{Nonmag1}
\figsetplot{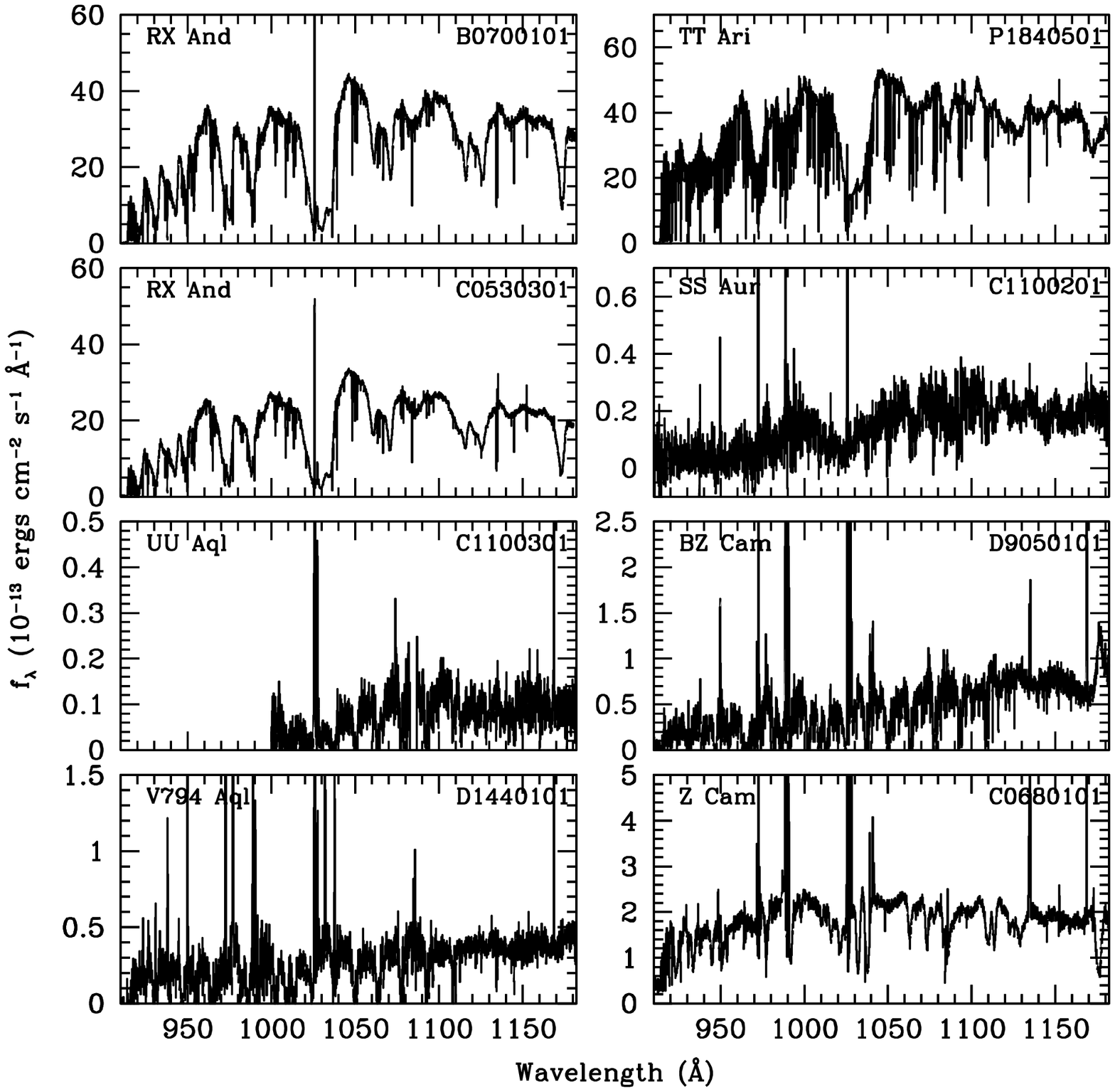}
\figsetgrpnote{Non-Magnetic CVs}
\figsetgrpend

\figsetgrpstart
\figsetgrpnum{1.2}
\figsetgrptitle{Nonmag2}
\figsetplot{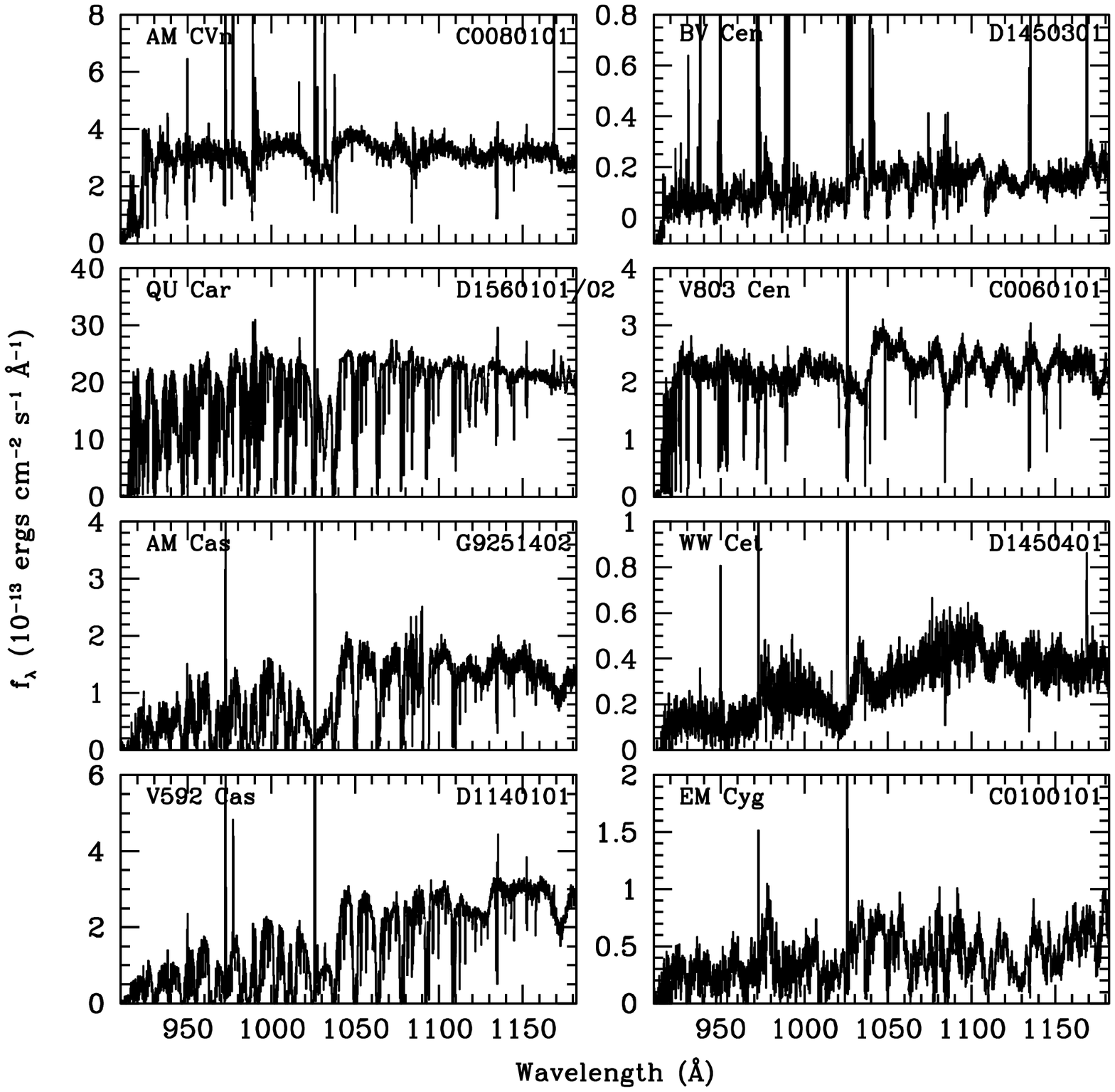}
\figsetgrpnote{Magnetic CVs}
\figsetgrpend

\figsetgrpstart
\figsetgrpnum{1.3}
\figsetgrptitle{Nonmag3}
\figsetplot{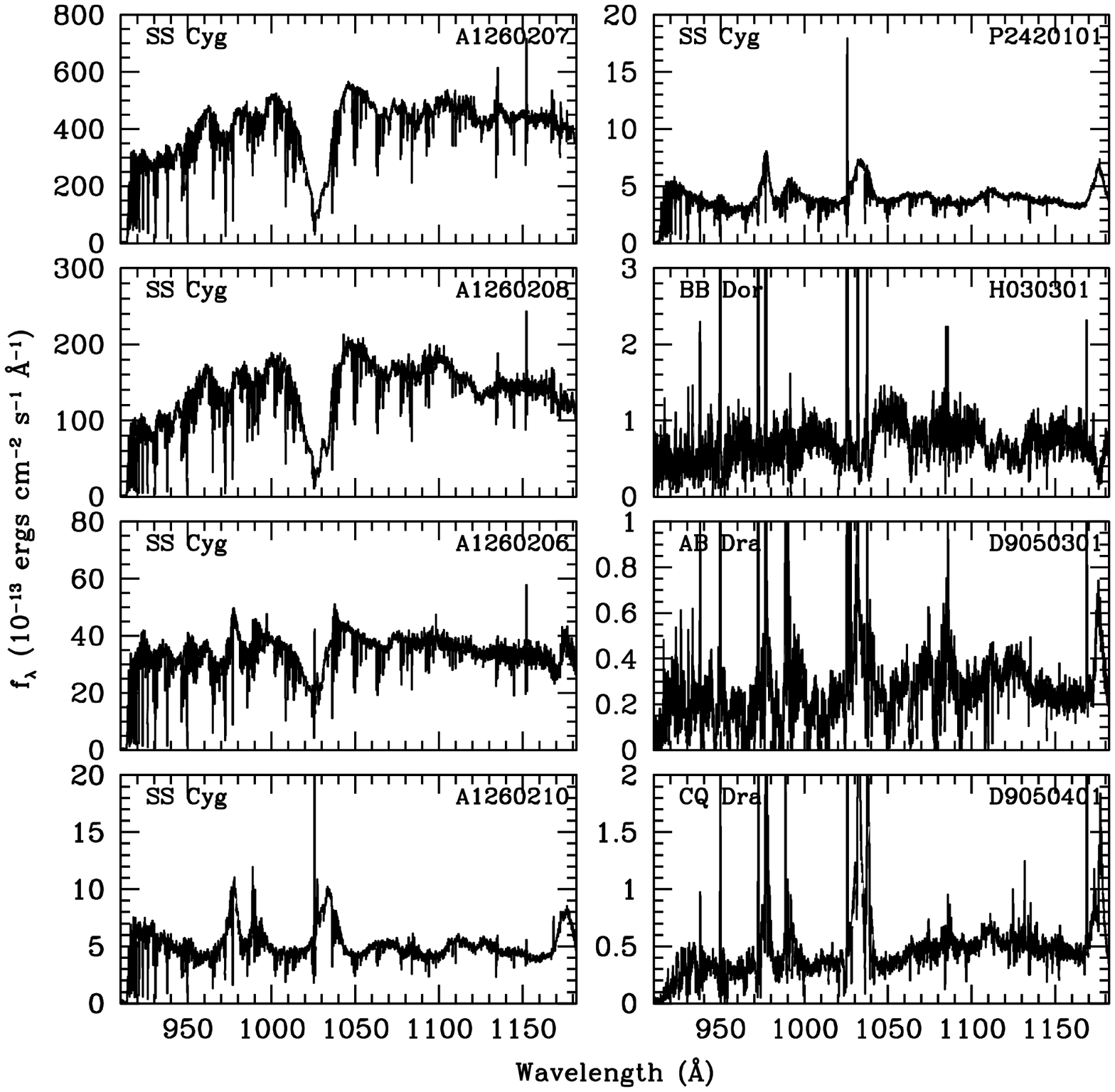}
\figsetgrpnote{Non-Magnetic CVs}
\figsetgrpend

\figsetgrpstart
\figsetgrpnum{1.4}
\figsetgrptitle{Nonmag4}
\figsetplot{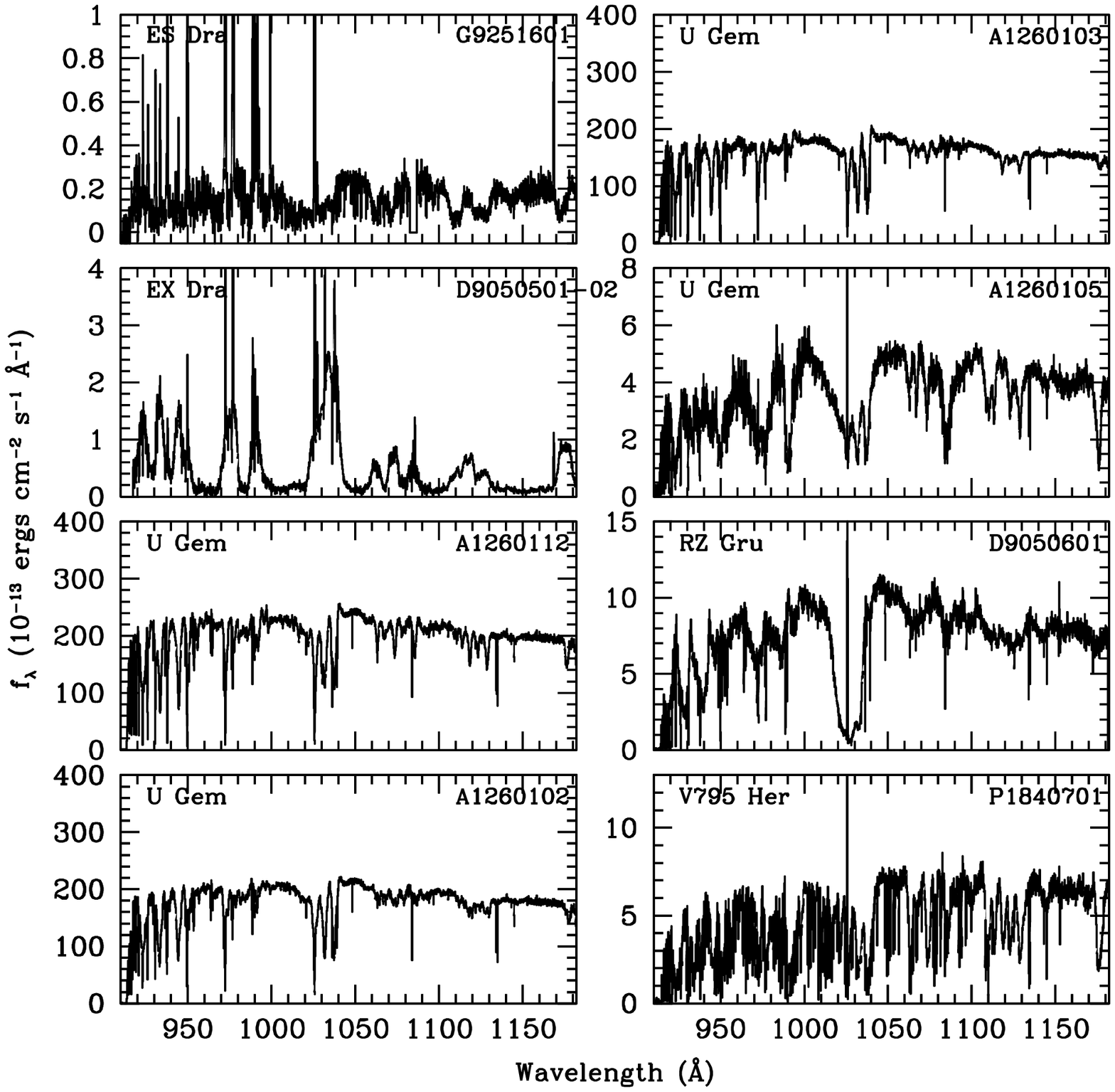}
\figsetgrpnote{Non-Magnetic CVs}
\figsetgrpend

\figsetgrpstart
\figsetgrpnum{1.5}
\figsetgrptitle{Nonmag5}
\figsetplot{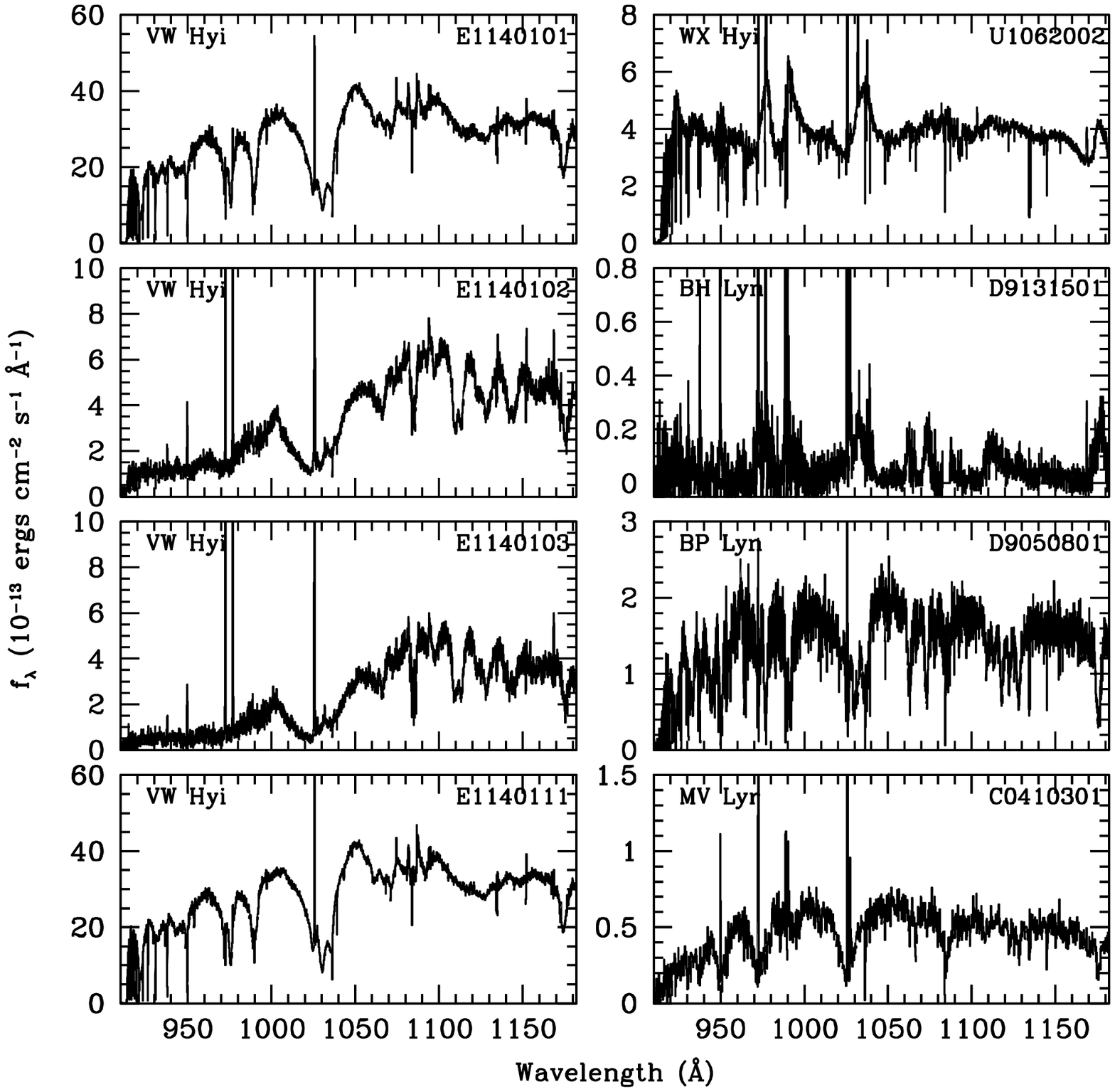}
\figsetgrpnote{Non-Magnetic CVs}
\figsetgrpend

\figsetgrpstart
\figsetgrpnum{1.6}
\figsetgrptitle{Nonmag6}
\figsetplot{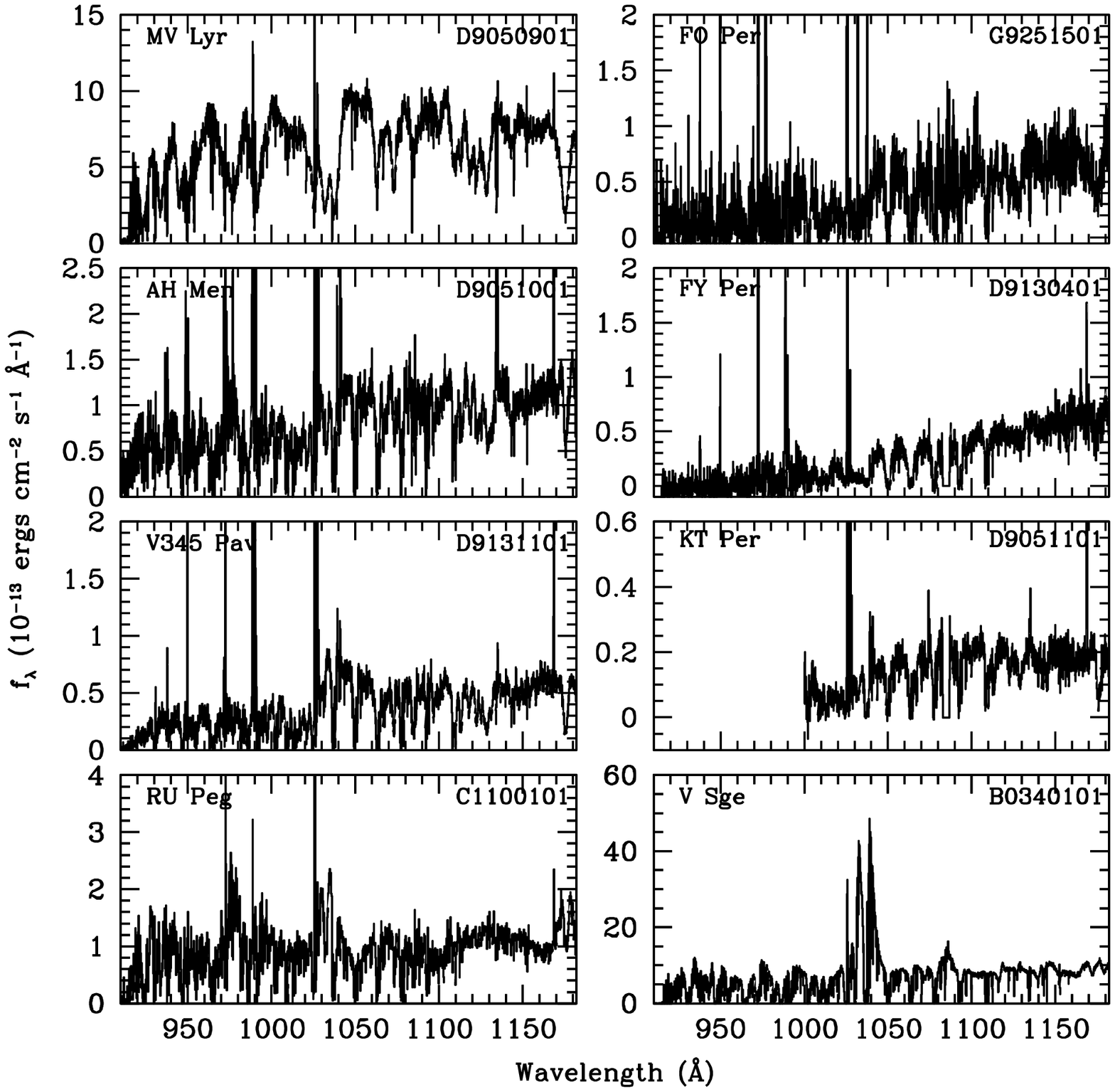}
\figsetgrpnote{Non-Magnetic CVs}
\figsetgrpend

\figsetgrpstart
\figsetgrpnum{1.7}
\figsetgrptitle{Nonmag7}
\figsetplot{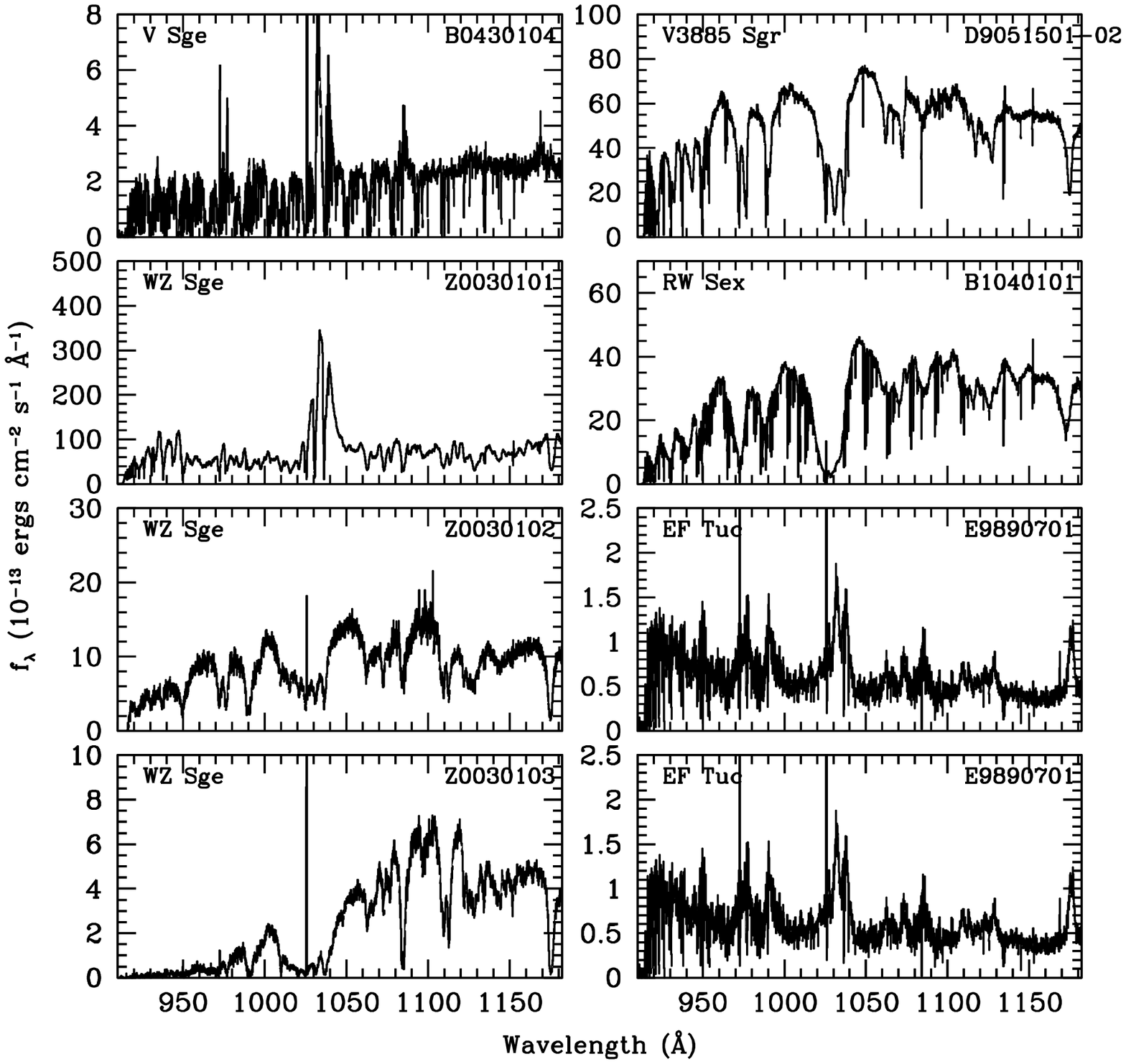}
\figsetgrpnote{Non-Magnetic CVs}
\figsetgrpend

\figsetgrpstart
\figsetgrpnum{1.8}
\figsetgrptitle{Nonmag8}
\figsetplot{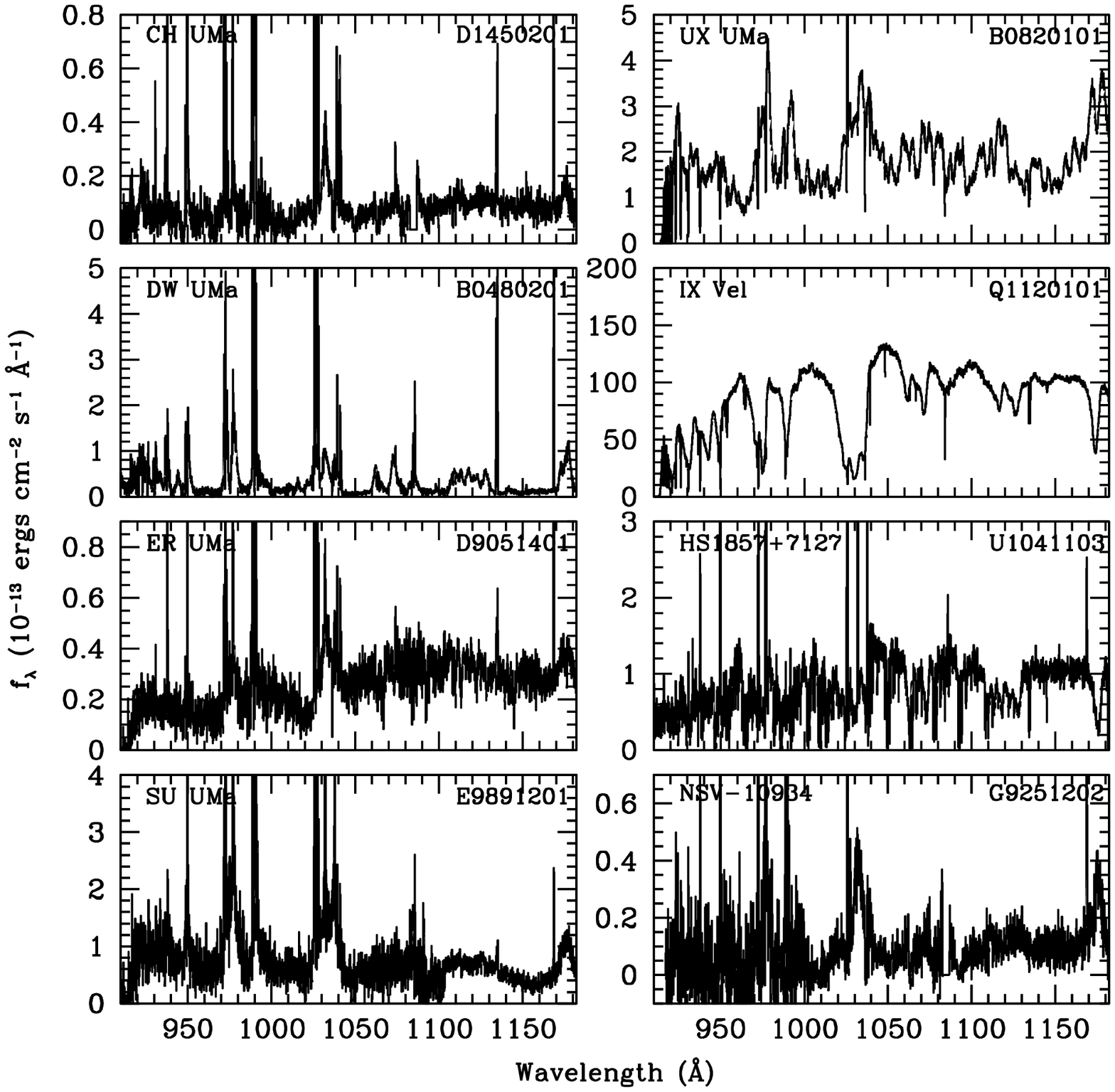}
\figsetgrpnote{Non-Magnetic CVs}
\figsetgrpend

\figsetgrpstart
\figsetgrpnum{1.9}
\figsetgrptitle{Nonmag9}
\figsetplot{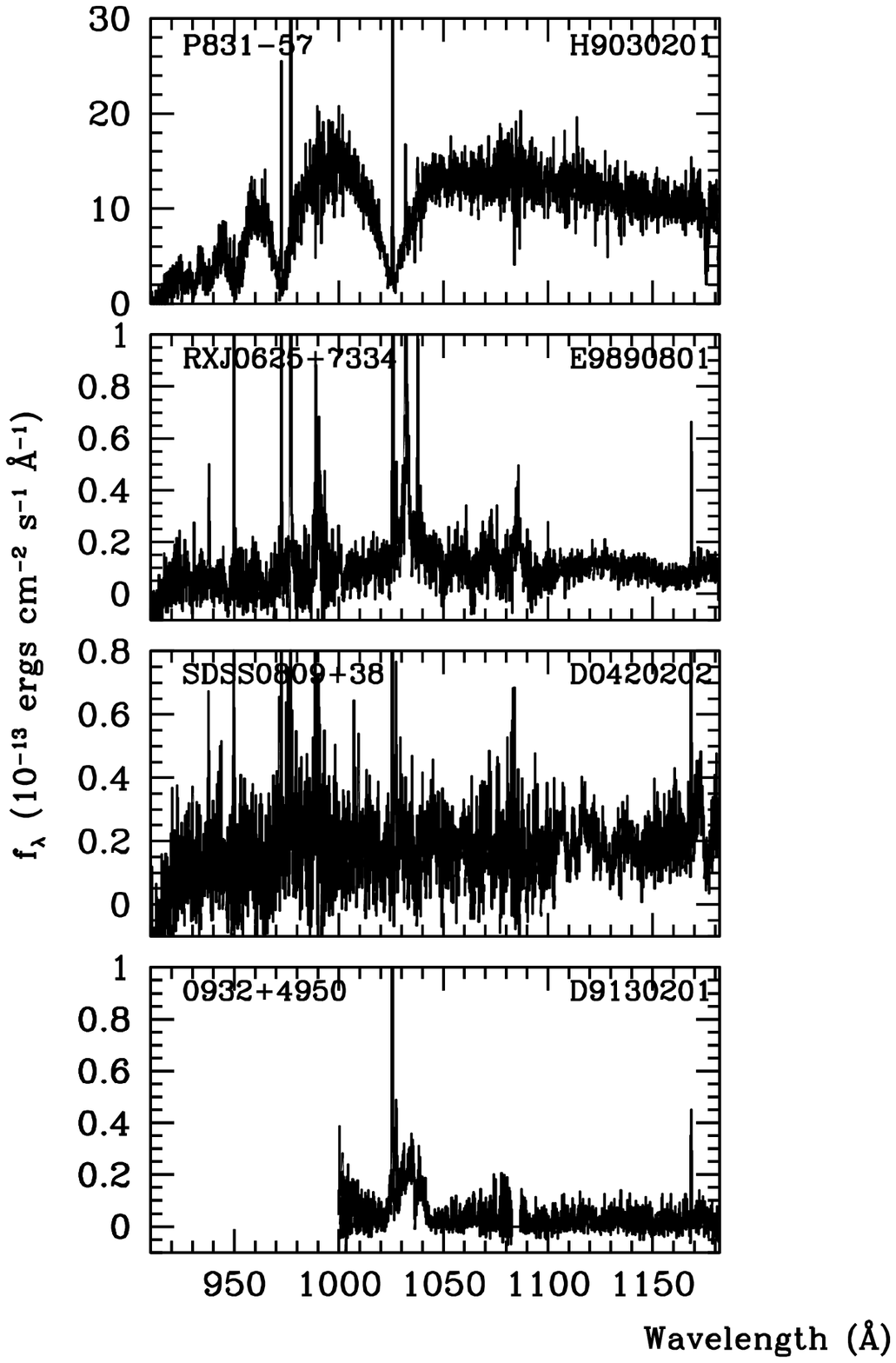}
\figsetgrpnote{Non-Magnetic CVs}
\figsetgrpend

\figsetend

\figsetstart
\figsetnum{2}
\figsettitle{Magnetic CVs Observed by \textit{FUSE}}

\figsetgrpstart
\figsetgrpnum{2.1}
\figsetgrptitle{Mag1}
\figsetplot{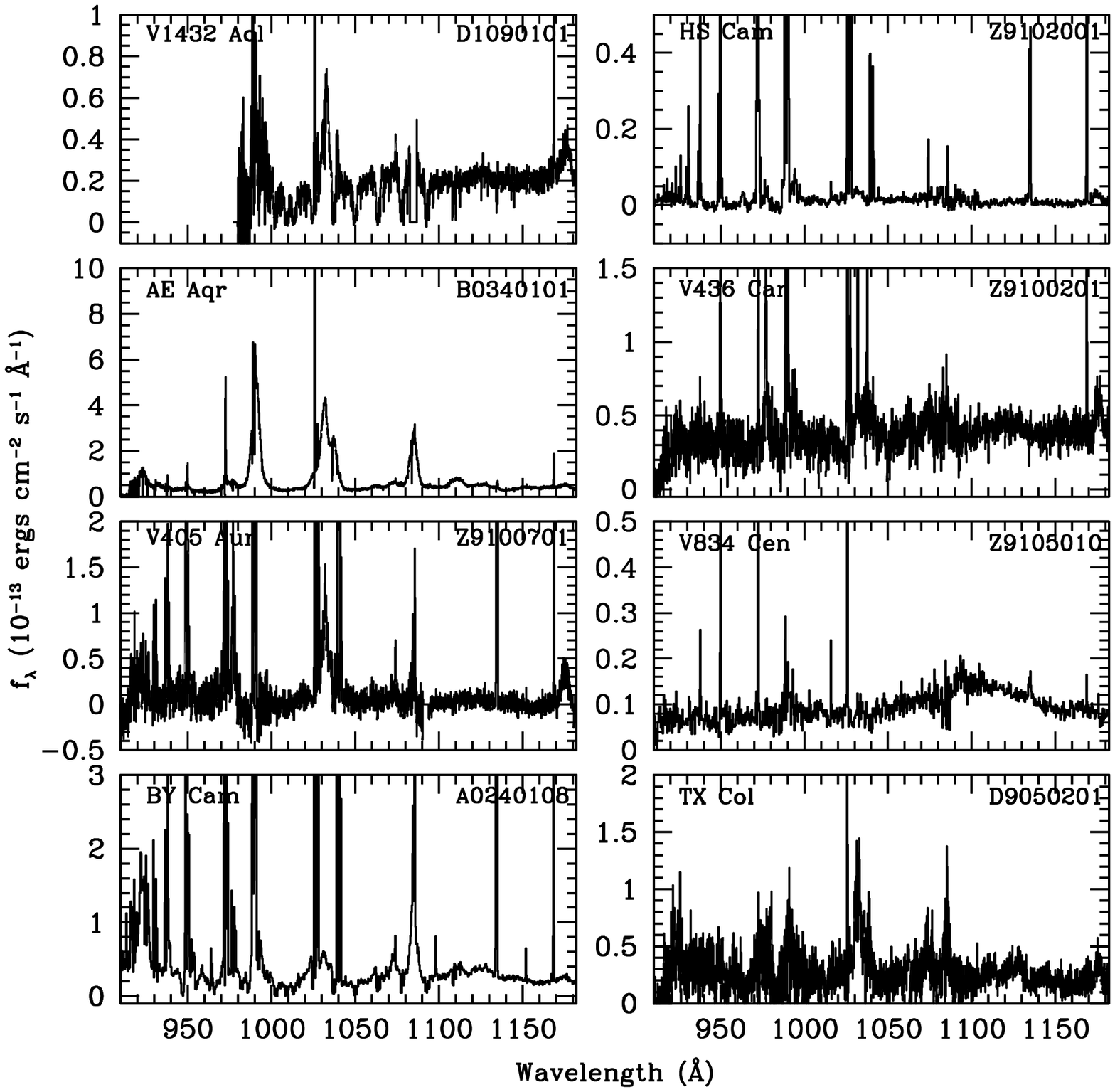}
\figsetgrpnote{Magnetic CVs}
\figsetgrpend

\figsetgrpstart
\figsetgrpnum{2.2}
\figsetgrptitle{Mag2}
\figsetplot{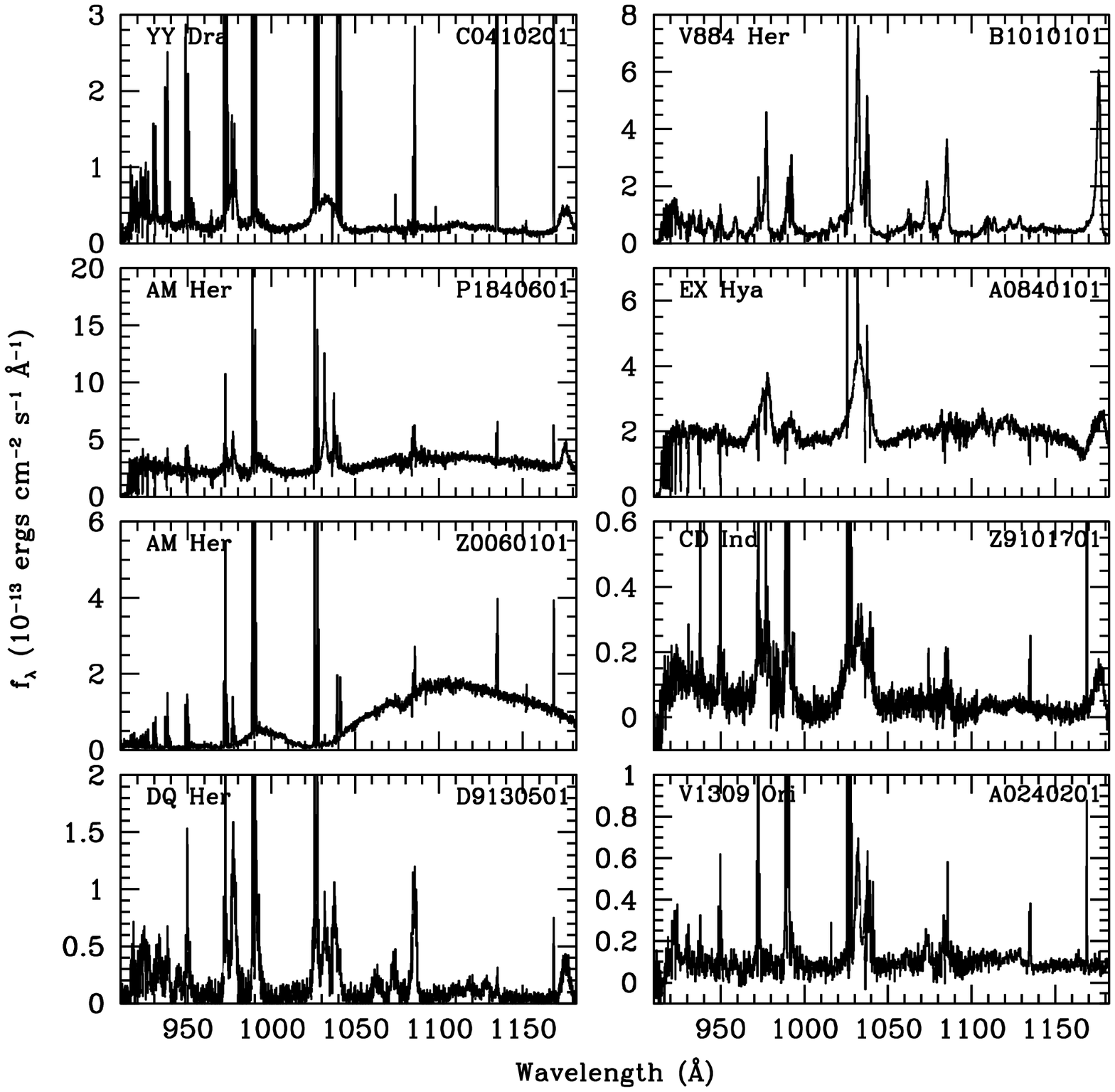}
\figsetgrpnote{Magnetic CVs}
\figsetgrpend

\figsetgrpstart
\figsetgrpnum{2.3}
\figsetgrptitle{Mag3}
\figsetplot{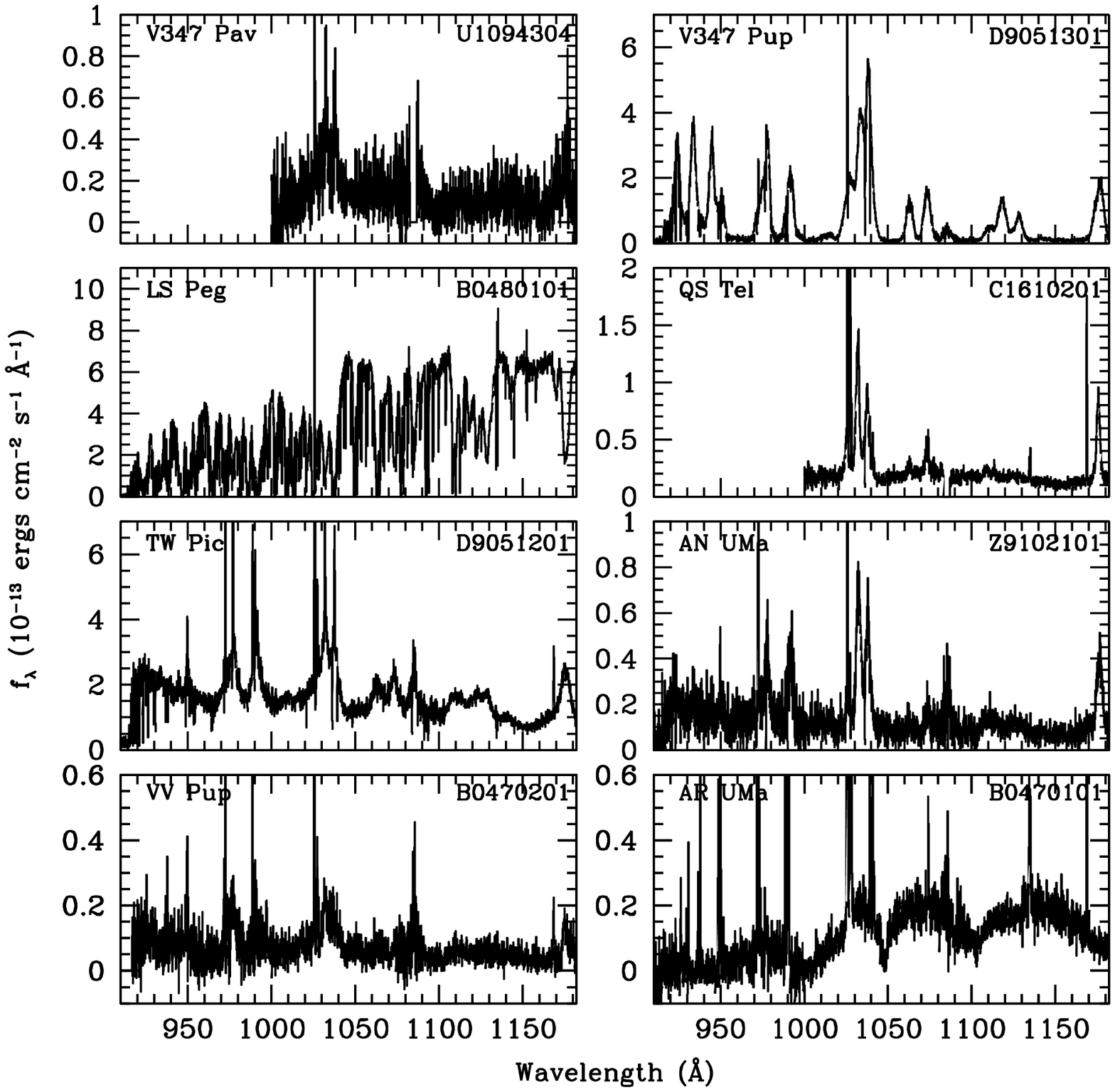}
\figsetgrpnote{Magnetic CVs}
\figsetgrpend

\figsetgrpstart
\figsetgrpnum{2.4}
\figsetgrptitle{Mag4}
\figsetplot{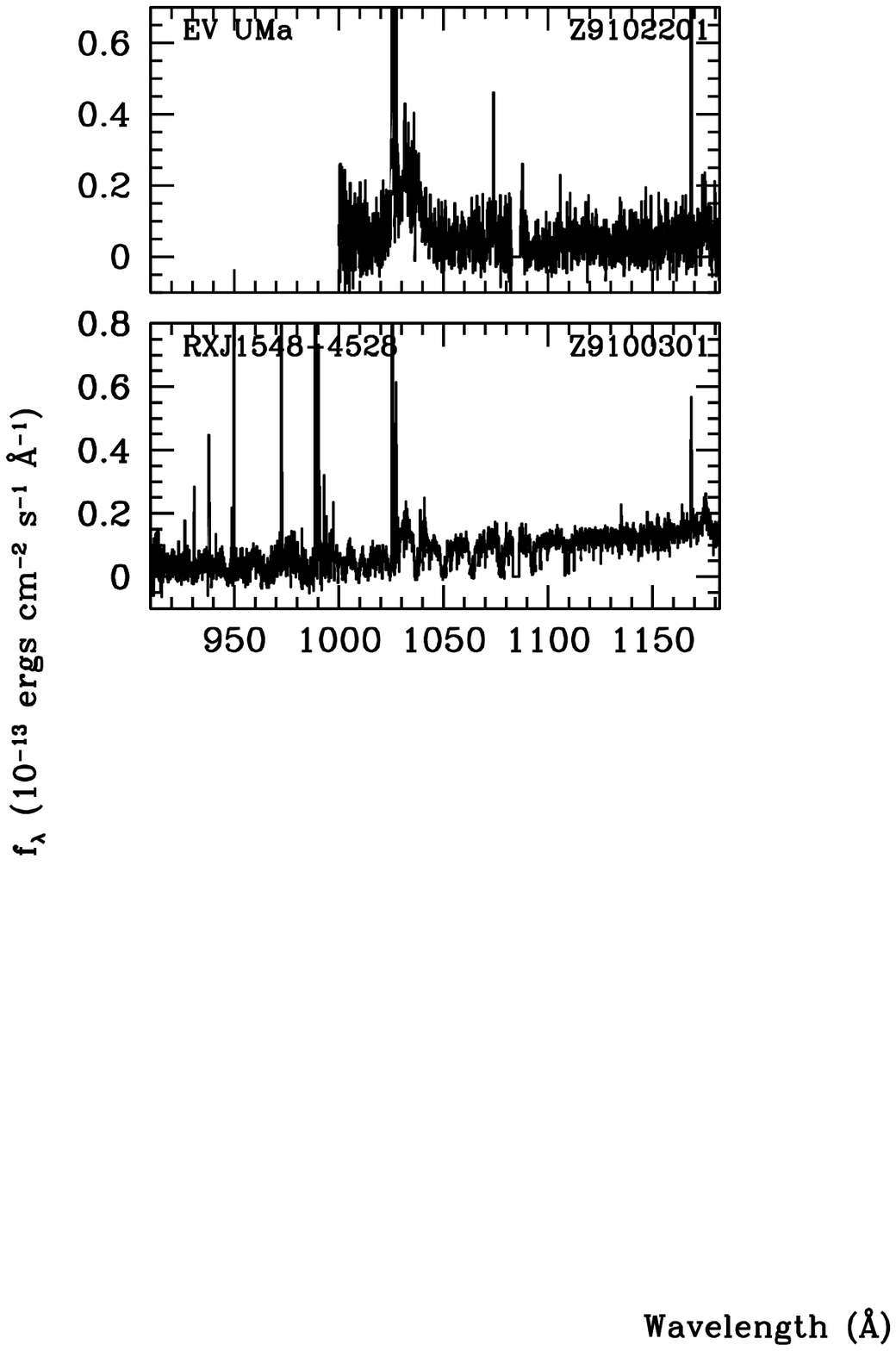}
\figsetgrpnote{Magnetic CVs}
\figsetgrpend
\figsetend

\clearpage
\pagestyle{empty}

\clearpage
\pagestyle{empty}

\clearpage
\pagestyle{empty}

\clearpage
\pagestyle{empty}

\clearpage
\pagestyle{empty}

\clearpage
\pagestyle{empty}

\clearpage
\pagestyle{empty}

\clearpage
\pagestyle{empty}

\clearpage
\pagestyle{empty}

\clearpage
\pagestyle{empty}

\end{document}